\documentclass[fleqn,usenatbib]{mnras}

\usepackage{newtxtext,newtxmath}

\usepackage[T1]{fontenc}
\usepackage{ae,aecompl}

\usepackage{graphicx}	
\usepackage{amsmath}	
\usepackage{appendix}
\usepackage{bm}

\title[Deep learning in galaxy formation]{Efficient exploration and calibration of a semi-analytical model of galaxy formation with deep learning}

\author[E. J. Elliott et al.]{
Edward J. Elliott,$^{1,2}$\thanks{E-mail: edward.j.elliott@durham.ac.uk }
Carlton M. Baugh,$^{1,2}$
Cedric G. Lacey$^{1}$
\\
$^{1}$Institute for Computational Cosmology, Department of Physics, Durham University, South Road, Durham DH1 3LE, UK.\\
$^{2}$Institute for Data Science, Department of Physics, Durham University, South Road, Durham DH1 3LE, UK.\\
}

\date{Accepted XXX. Received YYY; in original form ZZZ}

\pubyear{2021}
\begin{document}
\label{firstpage}
\pagerange{\pageref{firstpage}--\pageref{lastpage}}
\maketitle

\begin{abstract}
We implement a sample-efficient method for rapid and accurate emulation of semi-analytical galaxy formation models over a wide range of model outputs. We use ensembled deep learning algorithms to produce a fast emulator of an updated version of the \texttt{GALFORM} model from a small number of training examples. We use the emulator to explore the model's parameter space, and apply sensitivity analysis techniques to better understand the relative importance of the model parameters. We uncover key tensions between observational datasets by applying a heuristic weighting scheme in a Markov chain Monte Carlo framework and exploring the effects of requiring improved fits to certain datasets relative to others. Furthermore, we demonstrate that this method can be used to successfully calibrate the model parameters to a comprehensive list of observational constraints. In doing so, we re-discover previous \texttt{GALFORM} fits in an automatic and transparent way, and discover an improved fit by applying a heavier weighting to the fit to the metallicities of early-type galaxies. The deep learning emulator requires a fraction of the model evaluations needed in similar emulation approaches, achieving an out-of-sample mean absolute error at the knee of the K-band luminosity function of 0.06 dex with less than 1000 model evaluations. We demonstrate that this is an extremely efficient,  inexpensive and transparent way to explore multi-dimensional  parameter spaces, and can be applied more widely beyond semi-analytical galaxy formation models.

\end{abstract}




\section{Introduction}

Galaxy formation is a complex and non-linear process, involving the interplay of gravitational, radiative, thermal, and fluid processes. Semi-analytical modelling is an approach used to improve our understanding of this problem by reducing it to its key ingredients using simplified mathematical relations motivated by physical and geometric arguments \citep[e.g.][]{Baugh2006, Benson2010a}. These relations take the form of coupled differential equations and simple algebraic relations describing processes such as star formation, gas cooling, and bar instabilities in galactic disks. Semi-analytical models provide a comprehensive theoretical framework with which to understand and develop intuition about galaxy formation, and have produced a number of insights \cite[e.g.][]{White1991, Benson2003, Bower2006, Croton2006,  Lacey2016}.

However, the semi-analytical approach has sometimes attracted scepticism for a number of reasons. The mathematical relations which describe the physical processes in the model often contain adjustable parameters, and a model is defined by a particular choice for the parameters values \citep[analogous to the parametrised sub-grid models employed in hydrodynamic simulations, e.g.][]{Crain2015, Somerville2015} These parameters are sometimes set by theoretical or observational considerations, but in many cases they are less well specified (for example, in the case of the parameters governing the strength of feedback due to supernovae - SNe).

There is a perception--which we believe to be misplaced--that these `free' parameters allow semi-analytical models (SAMs) to fit any arbitrary combination of datasets, therefore eliminating their predictive and explanatory power. We hope to dispel this view by demonstrating that the majority of the variance in the model output is contributed by just a few parameters which have clear physical interpretations (such as the strength of feedback due to SNe or AGN), and that these dominant parameters preclude arbitrary fitting. 

Another major source of the scepticism towards SAMs arises from the seemingly opaque procedures that have  commonly been used to calibrate the model parameters. This process often follows a `chi-by-eye' methodology, in which the operator adjusts the parameters by hand, interprets the effect on the model output, and adjusts the parameters again to improve the match of the model output to an observable. This requires a high level of expertise and familiarity with the SAM, and the operator often makes trade-offs between fits to different constraining datasets in a way which is poorly defined; model predictions are often judged to be good fits when formally they would be rejected. This makes the process of setting the model parameters hard to reproduce. There is also no guarantee that the by-eye approach will produce the best fit to the calibration datasets; the model parameters may interact in a non-linear way, which can be difficult to conceptualize. This, coupled with the large parameter space, makes it unlikely that such a search will find the best-fitting parameters. We aim to side-step these issues by developing a method to rapidly and robustly perform an exhaustive search of the parameter space, calibrate the SAM in an automatic way without the need for significant human intervention, and quantify the relative importance of the parameters. In this way, we hope to make the model calibration process transparent and reproducible, especially by researchers with less experience of running SAMs. Although the cosmological parameters are now well constrained, SAMS must still be re-tuned for simulations with different resolutions and cosmologies, such as $f(R)$ gravity simulations, or when a new implementation of a process is included. The question of how to set the model parameters therefore remains a relevant one. 

The calibration and exploration of SAMs is not a new problem, and has been investigated in a number of previous works. This effort has generally taken two forms: direct exploration of the model parameter space, and emulation. Although SAMs are orders of magnitude cheaper than hydrodynamic simulations, direct exploration of their parameter space is computationally expensive due to the sheer number of model runs required for a formal search; often this will take a prohibitive length of time except for the case of tuning the parameters to a small number of datasets. This approach has been investigated in a number of papers. \cite{Kampakoglou2008} implemented Markov-Chain Monte Carlo (MCMC) techniques to calibrate a SAM to multiple datasets. \cite{Henriques2009} again used MCMC to calibrate the \texttt{L-GALAXIES} SAM to a number of a datasets, finding that the choice of datasets altered the values of  the best-fitting parameters, pointing to deficiencies in their model. \cite{Martindale2017} expanded on this to include the HI mass function as a constraint, leading to a change in the best-fitting parameters.  \cite{Lu2011, Lu2012} constrained the parameter space which gave acceptable fits to the K-band luminosity function (LF), and expanded this to include the HI mass function in \cite{Lu2014}. \cite{Ruiz2015} used particle swarm optimization to calibrate a SAM to the K-band LF. The second class of methods involves constructing a statistical emulator of the SAM which can be evaluated orders of magnitude more quickly than running the SAM itself, but at the cost of being approximate by nature.  \cite{Bower2010} and \cite{Vernon2010} employed a Bayesian emulation technique \citep[as developed by][]{bayeslinear} to constrain the parameter space which can provide reasonable fits to the K- and b$_{\rm J}$-band LFs, and extended this in \cite{Benson2010} to explore this ability of this reduced parameter space to fit to further observational datasets. This approach has also been applied by \cite{Rodrigues2017} to calibrate the \texttt{GALFORM} SAM to the galaxy stellar mass function in the local Universe, and recently by \cite{vandervelden2021} to calibrate the \texttt{Meraxes} SAM at high redshift.

Here, we aim to emulate an updated version of the \texttt{GALFORM} code implemented in the Planck Millenium N-body simulation \citep{Baugh2019}, which uses an improved galaxy merger scheme (devised by \citealp{Simha2017} and was first implemented in \texttt{GALFORM} by  \citealp{Campbell2015}), but which also includes an improved model for gas cooling in halos \citep[introduced by][]{Hou2018}. 

We focus specifically on using deep learning to build our emulator \citep[for an introductory review, see e.g.][]{Emmert-Streib2020}. This sub-field of machine learning uses stacked neural layers (hence \textit{deep}) to build flexible function approximators which are able to uncover non-linear relations in data without the need for a strongly pre-defined model, and have proven to be highly successful in astronomical applications \citep[e.g.][]{Ravanbakhsh2016, Schmit2018, Perraudin2019, He2019, Cranmer2019, zhang2019, deoliveira2020, Ntampaka2019}. We demonstrate that deep learning algorithms can be applied to accurately emulate SAMs over the full range of model outputs, and require a relatively small number of training examples to achieve good accuracy when compared to other methods. Since the deep learning emulator can be evaluated orders of magnitude faster than the time taken to run  \texttt{GALFORM}, we are able to run many simple MCMC chains to explore the parameter space, and investigate how calibration to different datasets constrains the model parameters. We achieve this by minimizing the absolute error between the emulator output and the data, and employing a heuristic weighting scheme to the different observational datasets to mimic the process employed by model practitioners. In this way, we hope to elucidate and automate the calibration process, as well as exhaustively search the parameter space of the model. 

This approach has a number of advantages over previous work. Non-emulation approaches such as MCMC and particle swarm optimization offer a powerful way to quantify parameter uncertainty and fit the model to a particular observable, but are limited in terms of exploring and understanding the full parameter space, and come at significant computational expense. Previous emulation approaches, though informative, also do not fully address our aims; they have focused on reducing the parameter space based on measures of implausibility (a measure which incorporates information about the emulator prediction and target data, and their variances, to rule out regions of parameter space). By iteratively refining more approximate emulators over a number of waves of model runs, these methods hone in on a region of parameter space which could plausibly contain good fits to a predefined set of just a few observables. Here, we focus on producing an emulator of the \texttt{GALFORM} model which is accurate across the entire parameter space. This allows us to explore the full parameter space of the model and fit to a wide range of observables, and to consider more diverse combinations of observables than has been attempted in previous work. We also aim to reduce the requirement for a large number of \texttt{GALFORM} evaluations. \cite{Rodrigues2017}, for example, used 7 waves of 5000 runs each to hone in on the region of parameter space which gave acceptable fits to the local galaxy stellar mass function; here, we limit ourselves to $<$ 1000 full \texttt{GALFORM} runs. In doing so we intend to develop a general method for investigating, understanding, and calibrating SAMs in an inexpensive, flexible, and reproducible way.

We also apply sensitivity analysis techniques to the model parameters, as recently applied to \texttt{GALFORM} by \cite{Oleskiewicz2019}. This allows us to judge the importance of different parameters by quantifying the proportion of the model variance due to a given parameter through sensitivity indices. We are also able gauge the degree of interaction between parameters, giving us important insight into the model. 

The layout of the paper is as follows. In \S~\ref{sec:theoretical_background} we review the theoretical background. We describe the key processes of \texttt{GALFORM} that are relevant to this work in \S~\ref{sec:GALFORM}. In \S~\ref{sec:emulator} we give a brief review of the deep learning approach and our emulator design, and in \S~\ref{sec:SAdescription} we give a description of the sensitivity analysis method. In \S\ref{sec:calibrationDatasetsDescription} we describe the observational constraints under consideration, and in \S~\ref{sec:parameterFittingDescription} we discuss how we find best-fitting parameters using MCMC. In \S~\ref{sec:results} we present our results. In \S\ref{sec:emulator_performance} we review the predictive performance of the emulator, in \S~\ref{sec:SAresults} we show the results of our sensitivity analysis and model exploration, and in \S~\ref{sec:calibration_tension_results} we present our model calibration results. In \S\ref{sec:Discussion} we discuss the merits of our methods and outline potential future work, and conclude in \S~\ref{sec:conclusion}.

\section{Theoretical background}
\label{sec:theoretical_background}

Here we briefly describe aspects of  \texttt{GALFORM} pertinent to this work (\S~\ref{sec:GALFORM}) and  describe the process of building a deep learning emulator and motivate our specific choice of model (\S~\ref{sec:emulator}). We then briefly describe sensitivity analysis (\S~\ref{sec:SAdescription}), the observational datasets considered (\S~\ref{sec:calibrationDatasetsDescription}), and our calibration scheme (\S~\ref{sec:parameterFittingDescription}).

\subsection{\texttt{GALFORM}}
\label{sec:GALFORM} 

\texttt{GALFORM} is a state-of-the-art \textit{ab initio} physically motivated semi-analytical model of galaxy formation. The model tracks the merger histories of dark matter haloes, the cooling of gas to form galactic disks, quiescent star formation in the disk, bursts of star formation associated with mergers or disk instabilities, the resultant feedback and gas ejection driven by supernovae, the role of heating by AGN in inhibiting gas cooling, and the chemical enrichment of stars and gas \citep[for a full description of \texttt{GALFORM}  see][]{Cole2002, Lacey2016}. Here we review some aspects of the code relevant to this work and the following discussion.

\subsubsection{Quiescent star formation in disks}

The model uses an empirical star formation law formulated by \cite{Blitz2006} \citep[and implemented in \texttt{GALFORM} in][]{Lagos2011} based on observations of nearby star-forming disk galaxies. The star formation rate in the disc is 

\begin{equation}
    \psi\textsubscript{disk} = \nu\textsubscript{SF}M\textsubscript{mol, disk}, 
\end{equation}

\noindent where $M\textsubscript{mol, disk}$ is the mass of molecular gas in the disk, and $\nu\textsubscript{SF}$ is a constant which we treat as an adjustable parameter within a reasonable range \citep{Bigiel2011}. The mass of molecular gas depends on the gas pressure in the mid-plane of the disc.  

\subsubsection{Supernova feedback}

Supernova feedback causes gas to be ejected from galaxies. The model assumes that this mass ejection is proportional to the instantaneous star formation rate, $\psi$, with a mass loading factor dependent on the circular velocity of the galaxy, $V_\textsubscript{c}$:

\begin{equation}
    \label{eq:SNfeedback}
    \dot{M}\textsubscript{eject} = \left(\frac{V\textsubscript{c}}{V\textsubscript{SN}}\right)^{-\gamma\textsubscript{SN}}\psi , 
\end{equation}

\noindent where both $V\textsubscript{SN}$ and $\gamma\textsubscript{SN}$ are model parameters. We can further distinguish   $V\textsubscript{SN}$ into $V\textsubscript{SN, disk}$ and $V\textsubscript{SN, burst}$, allowing for different values for feedback in quiescent star formation and bursts, although these parameters have generally been assumed to be equal in most previous versions of the model. Gas ejected from the halo is assumed to gradually return from a reservoir beyond the halo's virial radius to the hot gas reservoir at a rate given by

\begin{equation}
    \dot{M}\textsubscript{return} = \alpha\textsubscript{ret}\frac{M\textsubscript{res}}{\tau\textsubscript{dyn,halo}} ,
\end{equation}

\noindent where $\tau\textsubscript{dyn,halo}$ is the dynamical time of the halo, $M\textsubscript{res}$ is the mass in the reservoir beyond the virial radius, and $\alpha\textsubscript{ret}$ is a free parameter.

\subsubsection{Galaxy mergers}

In the model, galaxy mergers can trigger bursts of star formation and destroy galactic disks. We define two different thresholds, $f\textsubscript{ellip}$ and $f\textsubscript{burst}$. When a satellite galaxy with baryonic mass $M\textsubscript{b, sat}$ merges with a central galaxy with baryonic mass $M\textsubscript{b, cen}$ two types of mergers may occur. First, if $M\textsubscript{b, sat}/M\textsubscript{b, cen} \geqslant f\textsubscript{ellip}$ the merger is classified as a \textit{major} merger, and the disk component of the galaxy is destroyed and forms a spheroid. The cold gas in the disk is assumed to be consumed in a burst of star formation. Second, if $M\textsubscript{b, sat}/M\textsubscript{b, cen} < f\textsubscript{ellip}$, the merger is classified as \textit{minor}, and the disk survives the merger. In this case, the cold gas is consumed in a starburst if a second condition is met, $M\textsubscript{b, sat}/M\textsubscript{b, cen} \geqslant f\textsubscript{burst}$. Both $f\textsubscript{burst}$ and $f\textsubscript{ellip}$ are treated as free parameters. In the improved galaxy merger model of \cite{Simha2017}, once a subhalo can no longer be resolved, an analytic calculation of the merger time is made based on dynamical friction arguments. 

\subsubsection{Disk instabilities}

Galactic disks dominated by rotational motion can become unstable to bar formation if their degree of self-gravity is too large. The model follows the work of \citet{Efstathiou1982}, and assumes that disks become unstable if the criterion 

\begin{equation}
    \label{eq:diskinstab}
    \frac{V\textsubscript{c}(r\textsubscript{disk})}{(1.68GM\textsubscript{disk}/r\textsubscript{disk})^{1/2}} \le f\textsubscript{stab}
\end{equation}

\noindent is met, where $M\textsubscript{disk}$ is the total disk mass and $r\textsubscript{disk}$ is the disk half-mass radius. Numerical simulations of a suite of exponential stellar disks by \citet{Efstathiou1982} found a value of $f\textsubscript{stab} \approx 1.1$. while \citet{christodoulou1995} found a value of 0.9 for gaseous disks. A value of 0.61 or below corresponds to universally stable disks, since this is the value of the left hand side of Eqn.~\ref{eq:diskinstab} for a completely self-gravitating disk. We allow this parameter to vary within a reasonable range (see Table \ref{tab:ParamRanges}). We assume that unstable disks are disrupted by bar instabilities on a sub-resolution timescale such that all the mass is instantaneously transferred to the spheroid and any gas present takes part in a burst of star formation.

\subsubsection{SMBH growth and AGN feedback}

Supermassive black holes can inject energy into the halo gas, disrupting gas cooling. Hot halo accretion, BH-BH mergers, as well as starbursts can increase the mass of the black hole \citep{Bower2006, Griffin:2019}. In the case of starbursts, the mass accreted onto the SMBH is a fraction $f\textsubscript{SMBH}$ of the mass of stars formed, where $f\textsubscript{SMBH}$ is an adjustable parameter. AGN accretion is assumed to occur if both of the following conditions are met: (1) that the gas halo is in quasi-hydrostatic equilibrium, that is the condition

\begin{equation}
    \tau\textsubscript{cool}/\tau\textsubscript{ff} > 1/\alpha\textsubscript{cool} ,
\end{equation}

\noindent is met, where $\tau\textsubscript{cool}$ is the cooling time of the gas, $\tau\textsubscript{ff}$ the free-fall time, and $\alpha\textsubscript{cool}$ is an adjustable parameter; (2) The AGN power required to balance the radiative cooling luminosity is below a fraction $f\textsubscript{Edd}$ of the Eddington luminosity of the SMBH.

\subsection{Deep learning emulator}
\label{sec:emulator}

Before we consider observational data, we aim to construct a fast emulator of the \texttt{GALFORM} model using the \texttt{tensorflow}  deep learning  framework \citep{tensorflow2015-whitepaper}. We formulate the problem from the perspective of supervised learning. We treat the \texttt{GALFORM} model as an unknown function $\Hat{f}(\cdot)$ that takes some input vector $\mathbf{x}$, representing a set of values for the model parameters, and produces an output vector $\mathbf{y}$, representing one or many binned statistical properties of the resulting synthetic galaxy population (e.g. the values of the K-band luminosity function in given luminosity bins). Our goal is then to develop a fast and accurate approximation to the function $\hat{f}(\cdot)$ by training an emulator to predict $\textbf{y}$ given $\textbf{x}$.

Since \texttt{GALFORM} is computationally expensive to run (at least in comparison to a potential deep learning emulator), we are limited in how many evaluations of the code we can perform, and so limited in the number of input-output pairs, $(\mathbf{x}_{i}, \mathbf{y}_{i})$, we have to train our emulator. To sample the parameter space evenly and efficiently, we use Latin hypercube sampling \citep[as described in e.g.][]{Bower2010} to generate the model input parameters. This method aims to fill the target parameter space evenly given a fixed number of samples. After evaluating \texttt{GALFORM} at these points, we are therefore left with the pairs of vectors $(\mathbf{x}_{i}, \mathbf{y}_{i})$, corresponding to the input and output of the model. We separate the samples randomly into three sets: the training set, the validation set, and the holdout set. The training and validation sets will be used to train the emulator, and the holdout set will be kept separate so it can be used for evaluating the emulator's performance on out-of-sample data. The different roles of these sets are discussed further below. 

\begin{figure}
    \centering
    \includegraphics[width=\columnwidth]{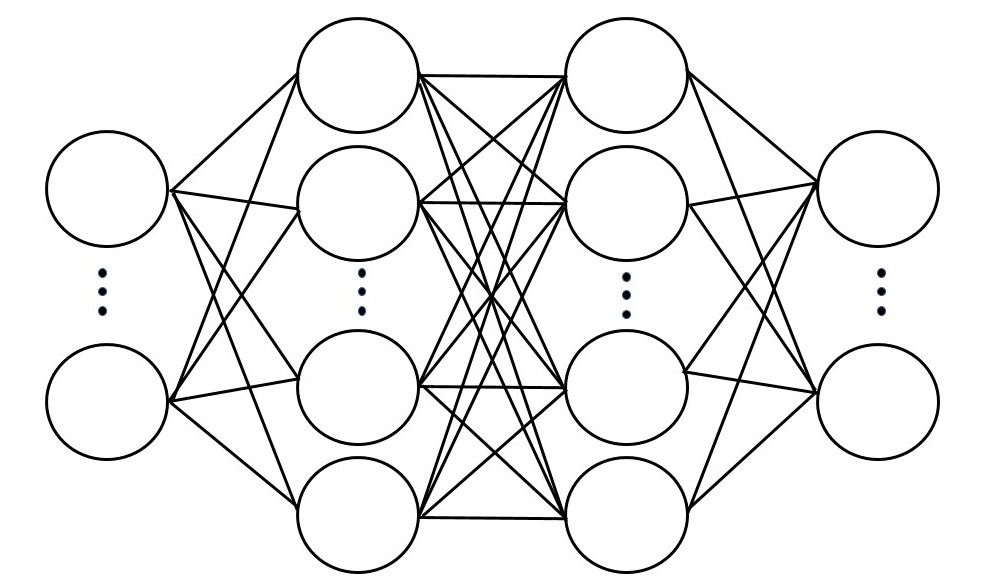}
    \caption{A schematic diagram showing a neural network with 2 hidden layers. The neurons on the left hand side represent the input layer, the central two layers of neurons are the hidden layers, and the right-hand neurons comprise the output layer.}
    \label{fig:ANN}
\end{figure}

The task of emulating \texttt{GALFORM} is therefore reduced to a regression problem. The deep learning emulator is comprised of stacked neural layers as shown in Fig.~\ref{fig:ANN}. Here we see a neural network with an input layer on the left, two \textit{hidden} layers, and an output layer on the right. Note that the output from each neuron is passed to every neuron in the following layer. The network is defined by a set of weights and biases, $W$; the $i$-th neuron in the $j$-th layer contains an adjustable weight vector $\mathbf{w}_{ij}$ and an adjustable bias term $b_{ij}$. When we  propagate inputs through our network to produce a prediction, the input layer first passes the inputs to every neuron in the first hidden layer. Each neuron $i$ in each subsequent layer $j$, starting with the first hidden layer, takes in the outputs from the previous layer and calculates its own output $z_{ij}$ by performing the computation

\begin{equation}
    z_{ij} = \hat{\sigma}(\textbf{z}_{j-1} \cdot \textbf{w}_{ij} + b_{ij}),
\end{equation}

\noindent where we have taken the dot product between the vector of all the neuron outputs in the previous layer $\mathbf{z}_{j-1}$ and the $i$-th neuron's weights $\mathbf{w}_{ij}$, and added the bias term $b_{ij}$. An activation function $\hat{\sigma}(\cdot)$ is then applied. This is generally a non-linear function such as the sigmoid or hyperbolic tangent function. The neuron outputs $z_{ij}$ are then passed to the next layer and the process is repeated until we reach the final layer. The output from the final layer is then the prediction of the network  for these inputs. Usually, the neurons in the final layer only apply a linear activation function. Therefore, since the network outputs are linear sums of non-linear functions of the input parameters, we can think of this method as estimating non-linear basis functions from training data. 

The weights and biases associated with each neuron are adjusted during training by seeking to minimise an error function between the emulator predictions and the true values. In our case, given a set of input parameters, we want to minimise the error between our emulator's prediction of the \texttt{GALFORM} output and the actual \texttt{GALFORM} output. We choose to use the mean absolute error function (hereafter MAE)

\begin{equation}
    \label{eq:MAE}
    \text{MAE} = \frac{1}{n}\sum_{k=1}^{n}{|\Hat{\textbf{y}}_{k} - \textbf{y}_{k}|},
\end{equation}

\noindent where $\Hat{\textbf{y}}_{k}$ is the model emulator prediction for the $k$-th of $n$ samples and $\textbf{y}_{k}$ is the true value. Since both $\Hat{\textbf{y}}_{k}$ and  $\textbf{y}_{k}$ are vector quantities, the modulus signs represent the L1 norm (i.e. the sum of absolute errors of the vector components);  we choose this metric as it gives less weight to outliers than the more commonly used L2 norm (i.e. the sum of squared errors of the vector components). If we denote the function represented by the neural network as $f$, parameterised by weights and biases $W$, we therefore attempt to find a function ${f}_{*}$ such that

\begin{equation}
f_{*} = \underset{W}{\operatorname{arg\:min}}  \{\text{MAE }(f(\mathbf{x}),\text{ }  \mathbf{y})\}.
\end{equation}

\noindent The training is performed iteratively in steps known as \textit{epochs}. During each epoch, the model weights and biases, $W$, are adjusted by an optimizer to minimise the MAE of the network's predictions for the training set. The optimizer is an algorithm which calculates how best to adjust the model weights by seeking minima on the error surface, usually by some form of gradient descent. We use the AMSGRAD variation of the Adam optimizer \citep{Kingma2015, AMSGRAD}. Adam is a momentum-based optimizer and AMSGRAD aims to improve the performance of Adam around minima on the error surface. At the end of each epoch, the adjusted model is evaluated on the validation set, to ensure the model generalises to unseen data. If the performance on the validation set has improved (as measured by the MAE), we save the model weights and continue training. If the performance does not improve, we do not save the weights and continue training. This process is repeated until the performance on the validation set has not improved for 30 epochs at which point we halt the training. We then perform a final fine-tuning step using the RMSprop optimizer \citep{RMSPROP}; this optimizer uses stochastic gradient descent and treats the error surface as a quadratic bowl. For this step, we use a very low learning rate of $10^{-5}$, allowing us to take small gradient-steps toward the minima of the error surface. We find this works well in boosting the performance of our emulator. We then evaluate the model on the holdout set to ensure its performance generalises to entirely unseen data (since we selected model weights which perform best on the validation set, the validation set itself is not a good test of out-of-sample performance).

\subsubsection{Inputs and outputs}

The aim of our emulator is to map an input vector $\mathbf{x}$, the \texttt{GALFORM} parameters, onto an output vector $\mathbf{y}$, the statistical galaxy properties that we wish to predict. Our choice of input parameters is informed by previous analyses \citep[e.g.][]{Lacey2016,Oleskiewicz2019}, and we aim to emulate the effects of the parameters associated with the key processes outlined in \S\ref{sec:GALFORM}.  These parameters and their ranges are shown in Table \ref{tab:ParamRanges}.
We train our emulator to predict a wide range of statistical galaxy properties calculated from the output of \texttt{GALFORM}. These are the K- and r-band LFs at $z = 0$, the early- and late-type galaxy sizes, the HI mass function, the early-type fraction with r-band magnitude, the I-band Tully-Fisher relation, the bulge-black hole mass relation, and the metallicities of early-type galaxies. 

\begin{table}
    \caption{The \texttt{GALFORM} parameters under investigation. See \S~\ref{sec:GALFORM} for the equations which define the symbols in the first column.}
    \begin{tabular}{l|l|l}
    \hline
         Parameter & Range & Process  \\ 
    \hline
         $f\textsubscript{stab}$ & $0.61-1.1$ & Disk instability \\
         $\alpha\textsubscript{cool}$ & $0.2-1.2$ & AGN feedback \\
         $\alpha\textsubscript{ret}$ & $0.2-1.2$ & SN feedback \\
         $\gamma\textsubscript{SN}$ & $1.0-4.0$ & SN feedback \\
         $V\textsubscript{SN, disk}$ $[\mathrm{kms}^{-1}]$& $100-550$ & SN feedback \\
         $V\textsubscript{SN, burst}$ $[\mathrm{kms}^{-1}]$ & $100-550$ & SN feedback \\
         $f\textsubscript{burst}$ & $0.01-0.3$ & Mergers \\
         $f\textsubscript{ellip}$ & $0.2-0.5$ & Mergers \\
         $\nu\textsubscript{SF}$ $[\mathrm{Gyr}^{-1}]$ & $0.2-1.7$ & Quiescent star formation \\
         $f\textsubscript{SMBH}$ & $0.001-0.05$ & SMBH accretion \\
         \hline
          
    \end{tabular}

    \label{tab:ParamRanges}
\end{table}

\subsubsection{Model architecture}

We find that a simple architecture is sufficient to accurately emulate  \texttt{GALFORM}. We use a densely-connected neural network, meaning that every neuron is connected to every neuron in the previous layer. We use two hidden layers, each with 512 neurons and sigmoid activation functions, and linear activations on the output layer. We investigated a number of other architectures, such as stacking long short term memory (LSTM;  \citealt{Hochreiter1997}) and 1D convolutional layers to try to exploit features of the data, but found limited improvement at the cost of slower evaluation speed. 

\subsubsection{Ensembling}

Training a neural network is a stochastic process. The network weights are often initialized according to some distribution \citep[e.g.][]{Glorot2010}, and the optimizer traverses the weight space using gradient steps calculated on mini-batches of the full dataset (i.e. a small subset of the whole training set at a time), and so is inherently stochastic. This means that training a single network is sub-optimal. Since the error surface is likely to contain many local minima we are unlikely to find the best possible network weights with one network alone, and each network will develop its own idiosyncrasies in how it fits the data. Neural networks also contain a vast number of parameters, and are therefore prone to over-fitting. One way to address these problems is ensembling. This involves training a  handful of networks with different weight initializations and combining the individual predictions. We can also shuffle the validation and training sets for each model in the ensemble, so that each model is exposed to a different distribution of input-output pairs. In general, this allows for a more robust prediction. Individual models may over- or under-fit to different features of the data, and combining predictions averages over these individual behaviours. 

We therefore train 10 models as described above, each with the same model architecture. Our emulator is then the simple average of the predictions of this ensemble of models. We must note however that this is a rich avenue for exploration in future work \citep[for a review of popular ensembling methods, see][]{Maclin2016}. For example, it may be possible to ensemble different machine learning algorithms and combine the individual model predictions with a weighting scheme, or even another machine learning algorithm.

\subsection{Sensitivity analysis}
\label{sec:SAdescription}
Once we have trained a deep learning emulator of \texttt{GALFORM}, we can apply sensitivity analysis techniques \citep[e.g.][]{Saltelli2010, Saltelli2017} to understand the contribution of the different parameters to the bin-wise variance in the emulator outputs. For a full description of calculating sensitivity indices see \cite{Oleskiewicz2019}, who first applied this type of analysis to a model of the entire galaxy population. Here we provide a brief overview of the sensitivity indices and what they describe. 

Since \texttt{GALFORM} is deterministic, all the variance in the output $Y$ will be due to the effects of the input parameters $X$. Assuming the input parameters are independent, we can calculate the first-order variance due to parameter $X_{i}$ by integrating the variance over the $i$-th dimension. Furthermore, we can calculate the variance due to interactions between parameters $X_{i}$ and $X_{j}$ by integrating the variance across the $i$-th and $j$-th dimensions, and subtracting the corresponding first-order effects of parameters $X_{i}$ and $X_{j}$. This can be continued to account for the interactions between many parameters. The total variance of the model output $Y$ can therefore be decomposed as

\begin{equation}
    \sum_{i=1}^{d}\text{Var}_{i} + \sum_{i<j}^{d}\text{Var}_{i,j} + ... + \text{Var}_{1,2...d} = \text{Var}(Y)
\end{equation}

\noindent where $\text{Var}_{i}$ represents the variance due to the $i$-th of the $d$ parameters, the sum over $\text{Var}_{i,j}$ represents the variance due to interactions between the parameters $X_{i}$ and $X_{j}$, and $\text{Var}(Y)$ is the total variance in the model output $Y$. This can be normalised to give the sensitivity indices of all orders

\begin{equation}
    \sum_{i=1}^{d}\text{S}_{i} + \sum_{i<j}^{d}\text{S}_{i,j} + ... + \text{S}_{1,2...d} = 1.
\end{equation}

\noindent This can be separated into $S_{1}$, the first order sensitivity index, which describes the proportion of the variance due to the $i$-th parameter, and $S_T$, which encapsulates the proportion of variance due to the $i$-th parameter and all higher order interactions between the $i$-th parameter and all other parameters.

Given the low computational cost of our emulator, we can evaluate it at a large number of points in the parameter space following Saltelli sampling. This sampling method aims to both evenly sample the space and minimise the model discrepancy \citep[a concept whose full explanation is beyond the scope of this work, but is described in][]{Saltelli2010}, allowing for sample-efficient calculation of the sensitivity indices. For this analysis, we use the \texttt{SALib} python package \citep{Herman2017}.

\subsection{Calibration and comparison datasets}
\label{sec:calibrationDatasetsDescription}

We will use our emulator to calibrate  \texttt{GALFORM} using a number of datasets. For the most part, we adopt the datasets used for model calibration in \cite{Lacey2016}, but with a focus on low-redshift observations. The key change we make is to the choice of LF data. We use the K- and r-band LFs from the GAMA survey \citep{Driver2012}; we choose these datasets as they correspond to the same survey volume and the same analysis methods are used for each band, with consistent $k$-corrections to $z = 0$ bands. The measured LFs should therefore be as consistent as possible, allowing our model to fit both. We apply a number of selection criteria to the \texttt{GALFORM} output to replicate the observational samples of the calibration datasets. 

The full list of calibration and comparison datasets and their respective selection criteria are:

\begin{enumerate}
  \item For the K-band LF, we calibrate to data from \cite{Driver2012} and also compare to data from \cite{Kochanek2001}.
 \item For the r-band LF, we calibrate to \cite{Driver2012}.
  \item For the early- and late-type sizes, we calibrate to data from \cite{Shen2003}. We define early types in the model as galaxies with bulge-to-total $r$-band luminosities of $(B/T)\textsubscript{r} > 0.5$ and late types as $(B/T)\textsubscript{r} < 0.5$. Since the  half-light radii of late-type galaxies are measured in circular apertures projected on the sky, the late-type galaxy sizes are corrected to face-on values by multiplying the median sizes by a factor of 1.34 \citep[as in][]{Lacey2016}.
  \item For the HI mass function, we calibrate to data from \cite{Zwaan2005} and compare to the estimate from \cite{Martin2010}.
  \item For the early-type fraction, we calibrate to data $(B/T)_{r}$ derived from \cite{Moffett2016} (A. Moffett, private communication). Here, the $(B/T)_{r}$ ratio was calculated from GAMA using the  disk/bulge decomposition method outlined in \cite{Lange2016}.  We also compare to data from \cite{Gonzalez2009}, which uses concentration indexes calculated from SDSS data \citep{York2000}.   Again, early types are defined to have $(B/T)\textsubscript{r} > 0.5$.
  \item For the I-band Tully-Fisher relation we compare to a subsample of Sb-Sd galaxies from the \cite{Mathewson1992} catalogue, as selected by \cite{deJong2000}. Model galaxies are selected with $(B/T)\textsubscript{B} < 0.2$ and gas fractions $M\textsubscript{cold}/M_{*} > 0.1$, where $M\textsubscript{cold}$ is the cold gas mass and $M_{*}$ is the stellar mass.
  \item For the Bulge-BH mass relation, we compare to data from \cite{Haring2004}. To match the bias toward early-types in the sample, we choose model galaxies with $(B/T)\textsubscript{r} > 0.3$.
  \item For the early-type metallicity, we compare to data from \cite{Smith2009}. We select model galaxies which reside in dark matter halos with $M\textsubscript{halo} > 10^{14}h^{-1}M_{\odot}$ and define early-types as before. The observed metallicities are corrected for metallicity gradients as described in \cite{Lacey2016}.
  \item Finally, we explore the model predictions for data in a very different redshift range to our calibration datasets. We test the calibrated model predictions against observational estimates of the star formation rate density (SFRD) with redshift. We compare to data from \cite{Burgarella2013, Cucciati2012, Oesch2013, Sobral2013} and \cite{Gunawardhana2013}. Since the observationally derived SFR values depend on an assumed initial mass function, and our model assumes a mildly top-heavy initial mass function in starbursts, we account for this in the observational comparison by applying an approximate correction in which we weight the starburst SFR by a factor of 1.9 (see \cite{Lacey2016} for further details).
\end{enumerate}

\subsection{Parameter fitting}
\label{sec:parameterFittingDescription}

Once we have trained our emulator, we use Markov-Chain Monte Carlo (MCMC) to explore the effect of calibration to different datasets with a simple implementation of the Metropolis-Hastings algorithm \citep[e.g.][]{robert2016}. The complication here is that the observational errors on the datasets cannot be combined straightforwardly. For example, if we aimed to minimise $\chi^{2}$, and the error bar on a particular data point  in the constraining observational dataset was very small, this point would dominate the total error measure. Our MCMC chain would simply be trying to find the best fit to this one data point, without fitting to the others. We therefore aim to minimise the absolute error between the emulator output and the observational constraints, without considering the observational errors. This allows us to combine and fit to multiple datasets, without having to worry about the robustness of the associated observational error bars, and hence to avoid the complications described above. 

We also wish to have the flexibility to give more consideration to a selected observational constraint  over the others. This will allow us to investigate the effect of requiring better fits to some datasets, and to see how this affects the fit to other datasets, as well as how the optimal parameter choices change as a result. We therefore include a vector of heuristic weights, $\textbf{W}$, which can be varied to increase the contribution of the residuals from one constraint to the total error,

\begin{equation}
    \label{eq:MCMC}
    \text{MAE\textsubscript{obs}} = \frac{1}{n} \sum_{i=1}^{n}\frac{1}{n^{\text{obs}}_{i}}W_{i}|\Hat{\textbf{y}}_{i}- \textbf{y}^{\text{obs}}_{i}| ,
\end{equation}

\noindent where the sum is over the $n$ observational constraints, $W_{i}$ represents the weighting of the contribution to the total error of the $i$-th constraint, $\Hat{\mathbf{y}}_{i}$ represents the emulator prediction for a set of model parameters, and $\mathbf{y}^{\text{obs}}_{i}$ is the observational data for the $i$-th constraint with $n^{\text{obs}}_{i}$ datapoints. Since $\Hat{\mathbf{y}}_{i}$ and $\mathbf{y}^{\text{obs}}_{i}$ are vector quantities, the modulus signs represent the L1 norm. As the constraining datasets have a variety of values, we scale each one by a constant factor and apply a constant offset so that the range of each  $\mathbf{y}^{\text{obs}}_{i}$ is [0,1]. We apply the same scaling to the emulator predictions $\Hat{\mathbf{y}}_{i}$ before calculating Eqns.~\ref{eq:MAE} and \ref{eq:MCMC}. Note that since different datasets contain different numbers of datapoints, we divide the $i$-th dataset's error by the number of datapoints $n^\text{obs}_i$ so that each contributes equivalently to the mean error. In later sections when considering observational data, we shall refer to Eqn.~\ref{eq:MCMC} as just the mean absolute error (MAE).

We implement the Metropolis-Hastings algorithm as follows: we initialize each chain at a (different) random point in the parameter space, $\mathbf{x}$. We then draw the next sample in the chain, $\mathbf{x}'$, from independent Laplacian distributions, $\mathcal{L}(x_{i}' | \mu_{i}, {b_{i}}) =  \frac{1}{2b_{i}}\text{exp}(-|x_{i}'-\mu_{i}|/b_{i})$ with $\mu_{i}=x_{i}$ and the scale parameter for the $i$-th model parameter, $b_{i}$, taken to be 1/20th of the parameter ranges given in Table~\ref{tab:ParamRanges}. We then calculate the \textit{acceptance ratio}, $\alpha$, by taking the likelihood ratio of the emulator predictions to the observational data for the parameter sets $\mathbf{x}$ and $\mathbf{x}'$ under a Laplacian likelihood with scale parameter $b\textsubscript{obs} = 1/20$ (i.e. the ratio $\mathcal{L}(f_{*}(\mathbf{x}') | \bm{\mu},b\textsubscript{obs})/\mathcal{L}(f_{*}(\mathbf{x}) | \bm{\mu},b\textsubscript{obs})$, where $\bm{\mu}$ represents the values of the observational data and $f_{*}(\cdot)$ the emulator, and recalling we are using the modified absolute difference given in Eqn.~\ref{eq:MCMC}). We next generate a uniform random number $u\in[0,1]$; samples are  \textit{accepted} if $u \leq \alpha $, in which case we draw the next sample from Laplacians centered on $\mathbf{x}'$, or \textit{rejected} if $u > \alpha$, in which case we draw the next sample from Laplacians centered on the original point $\mathbf{x}$. Therefore, if the error between the emulator predictions for the parameter set $\textbf{x}'$ and the observational data is less than or equal to the error for the predictions for $\textbf{x}$, we accept the sample. If the error for $\textbf{x}'$ is not an improvement over the previous sample, we accept it with probability $\alpha$. The density of accepted samples should then trace the regions in the parameter space which give the best fits to the observational data. We discard the first 50\% of accepted samples to allow for burn-in. We test a number of values of the sampling Laplacian widths $b_{i}$ in the range $0.05-0.2$, in conjunction with the likelihood width $b\textsubscript{obs}$, and find that these parameters have little effect on the convergence of the chains, and larger $b\textsubscript{obs}$ simply increases the proportion of accepted samples. We ran longer chains up to 100,000 samples and found that they quickly converged to their minimum MAE (as given by Eqn.~\ref{eq:MCMC}) within the first 10,000 samples, and so choose this as our chain length.

\begin{figure*}
    \centering
    \includegraphics[width=\textwidth,trim={1cm 0.5cm 0.5cm 0cm},clip]{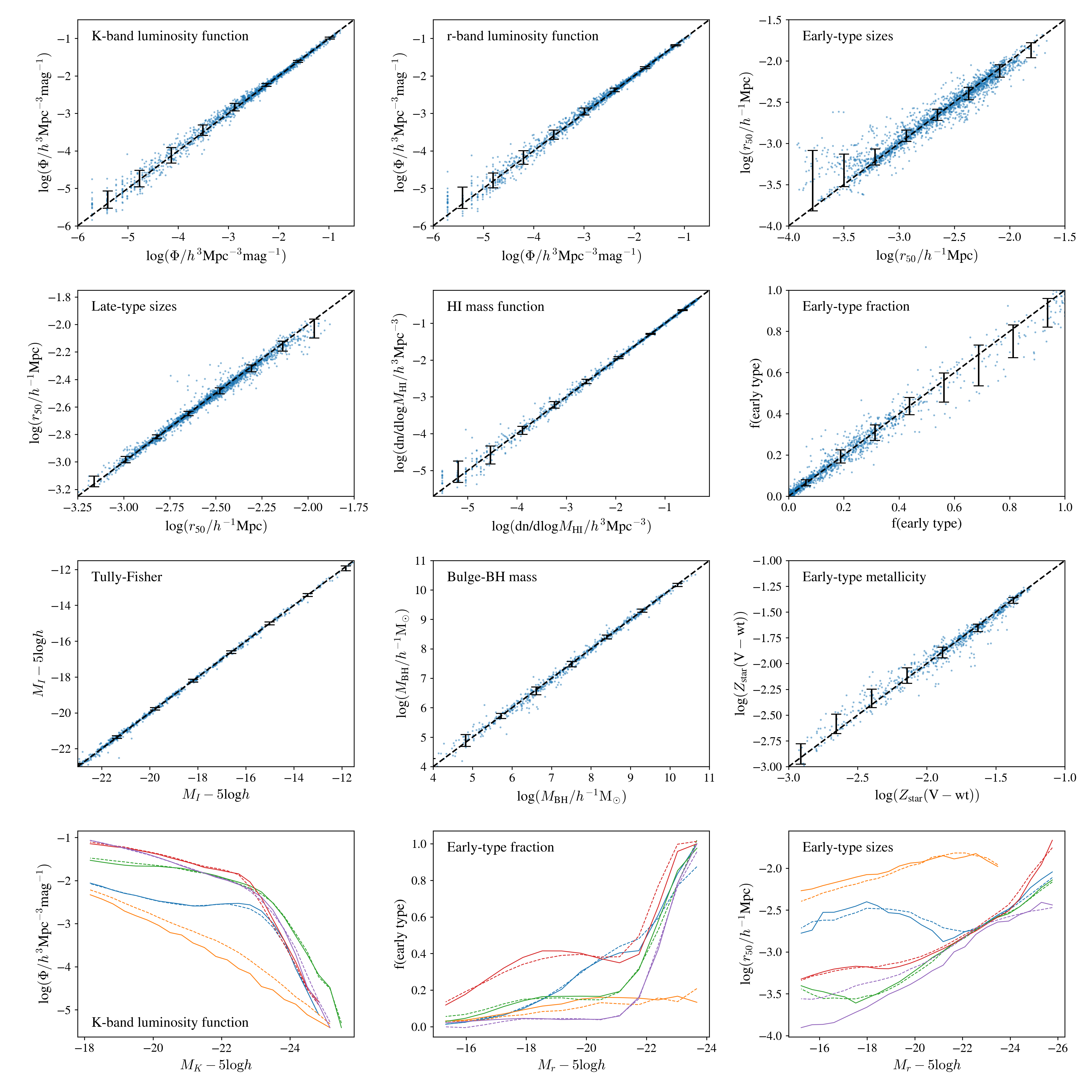}
    \caption{Emulator performance across nine statistics computed from the model output for out-of-sample parameter sets. These statistics are either number densities or median values in luminosity or mass bins, and are the same ones used for the observational comparisons. The first three rows show the emulator output (\textit{y}-axis) vs. the \texttt{GALFORM} output (\textit{x}-\text{axis}). Black error bars indicate the 10-90th percentile range of the residuals. The bottom row shows a random draw of emulator outputs (\textit{dotted}) and true \texttt{GALFORM} outputs (\textit{solid}) for the K-band LF, early-type fraction, and early-type sizes, reading from left to right. In these panels different colours denote different parameter sets.
    }
    \label{fig:EmulatorPerformance}
\end{figure*}

\section{Results}
\label{sec:results}

Here we present our main results, starting with a demonstration of the accuracy of the emulator (\S~\ref{sec:emulator_performance}), a sensitivity analysis of the model parameters (\S~\ref{sec:SAresults}), and 
closing with a discussion of the calibration of the model parameters and the tensions that arise when using different combinations of datasets  (\S~\ref{sec:calibration_tension_results}).

\subsection{Emulator performance}
\label{sec:emulator_performance}

Having trained our emulator as described in \S~\ref{sec:emulator}, we evaluate its ability to predict the  output of \texttt{GALFORM} at unseen points in the parameter space. We use a set of 930 \texttt{GALFORM} runs. The emulator was trained as described in \S~\ref{sec:emulator} with 80\% of the runs used as the training set (i.e. 744 combinations of parameter values), a 93 sample validation set, and a 93 sample holdout set. For each model in the emulator ensemble (i.e. each version of the neural network), the training and validation sets were shuffled. Fig.~\ref{fig:EmulatorPerformance} shows the emulator prediction vs. the true \texttt{GALFORM} output for the holdout set. Generally, the emulator follows a tight relation on the $y=x$ line, indicating that the emulator is accurately predicting the \texttt{GALFORM} output for the parameters sets in the holdout set. The HI mass function, Tully-Fisher relation, and Bulge-BH mass relations are accurately predicted, as well as the faint end of the luminosity functions and late-type galaxy sizes. The uncertainty is greater for the predictions of the bright-end of the LFs, and for the early-type sizes, fraction of early-type galaxies with luminosity, and the early-type metallicity. The lower panel of Fig.~\ref{fig:EmulatorPerformance} sheds some light on the source of inaccuracies in the early-type predictions, notably the early-type sizes, which exhibit noisy behaviour for some choices of parameters, and for a few cases (e.g. the purple line) the lower luminosity sizes are not well predicted. For the early-type fraction, while the error bars look large, inspection of the lower panels shows that such errors are generally in the brighter bins. We are nevertheless able to discriminate between parameter sets at the fainter magnitudes as the overall shape is well captured.

We see that the emulator is able to characterise a wide range of behaviour in the LFs, with the majority tightly predicted. In the bottom row of Fig.~\ref{fig:EmulatorPerformance}, the orange curves in the K-band panel show a substantial discrepancy between the true and predicted outputs; this usually indicates that the training data did not contain sufficient examples of this behaviour. The emulator constructs the function $f_{*}(\cdot)$ by fitting to the training examples, and in doing so should build a function which can interpolate between points in the parameter space. However, in sparsely sampled regions of the space, such as at the edges of our parameter bounds, the interpolation is less reliable. Indeed, if a point in the holdout set is an extrapolation with respect to the training set, performance can be affected. This is why we aim to fill the parameter space as evenly as possible using the Latin hypercube sampling method. We expect that such disagreements will decrease on increasing the number of training samples.

We can also judge from the distribution of predictions for the K- and r-band LFs in Fig.~\ref{fig:EmulatorPerformance} that the emulator slightly over-predicts the bright end of the LF. This is a consequence of the emulator training; in the interest of computing speed, we run \texttt{GALFORM} on only a sub-region accounting for 1\% of the full volume of the P-MILL simulation. This leads to sampling effects at low galaxy number densities, and for different choices of parameters the LF is cut off at different luminosities. Since the output of our emulator must be fixed-length, during training we mask any points beyond this luminosity cut-off when computing the loss. This means that in the brighter luminosity bins the emulator is only fitting to a small number of runs which are biased towards having higher values of $\phi$ in these brighter bins. There is therefore a tendency to over-predict at these luminosities. This should only be a minor problem in terms of our fitting routine, since the \cite{Driver2012} data we are fitting to does not sample $\phi$ to very low number densities. We also see a quantisation effect in the brighter LF bins, again due to the discrete sampling of galaxies. These problems could be removed by evaluating \texttt{GALFORM} on a larger fraction of the P-Mill simulation volume, though this would be more computationally expensive. 

\subsubsection{Scaling with training set size}

\begin{figure}
    \centering
    \includegraphics[width=\columnwidth,trim={0.4cm 0.75cm 0 0.5cm},clip]{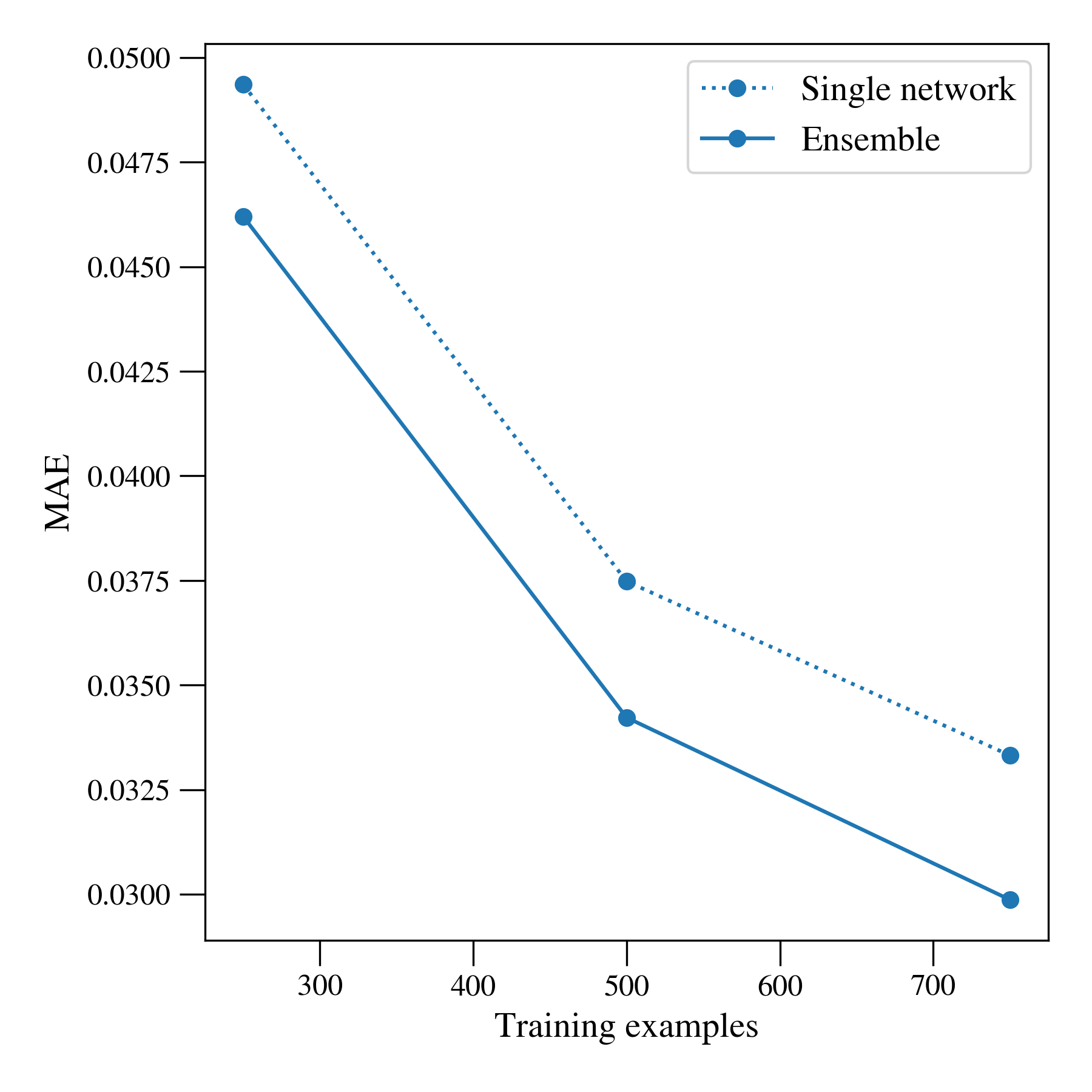}
    \caption{Emulator mean absolute error with the number of training examples of full \texttt{GALFORM} runs for the ensemble (solid line) and single (dotted line) networks. The emulators were trained on 250, 500 and 750 samples and performance evaluated on the same holdout set of 93 samples. Recall that the emulator outputs are scaled as described in \S~\ref{sec:parameterFittingDescription}.}
    \label{fig:EmulatorPerformanceSamples}
\end{figure}

We train three emulators with 250, 500 and 750 samples of parameters respectively (split into training and validation sets with 10\% of the samples being used for validation) to investigate the scaling of the emulator performance with the number of full \texttt{GALFORM} calculations carried out. The emulators consist of an ensemble of 10 networks each trained on the same (shuffled) training and validation data and the same holdout set of 93 samples. Fig.~\ref{fig:EmulatorPerformanceSamples} shows the scaling of the emulator performance on the holdout set (as measured by the MAE) with the number of training samples $N$. The dashed line shows average performance of the individual networks, and the solid line shows the performance of the ensemble. The model scales well with increasing training samples, and ensembling affords an almost constant improvement in performance (we find that at $\sim10$ models, the performance increase from adding more models to the ensemble saturates).

We test the ability of the emulator to generalise to unseen data by evaluating the version of the emulator trained with 500 samples in Fig.~\ref{fig:EmulatorPerformanceSamples} on the remaining 430 unseen samples. We find very little variation in the accuracy of the model between the two holdout sets. The MAE on the 93 sample holdout set was 0.034, and on the full 430 available holdout samples was 0.032. This gives us confidence that the model is able to learn a function which provides a very good approximation to \texttt{GALFORM}  across the full parameter space.

The impressive scaling of the emulator error with number of training samples is encouraging. SAMs are used to build mock catalogues for upcoming surveys, and some of these have stringent requirements on fits to certain datasets, such as the redshift distribution of galaxies. We can envisage using this technique to produce high accuracy parameter estimates for fits to such datasets by increasing the number of training samples, or using `zoom-in' training samples as in previous work  \citep[e.g.][]{Bower2010} to focus in on a particular region of parameter space which is deemed to give acceptable fits to the constraining datasets. Nevertheless, we find that our current emulator is sufficiently accurate to facilitate calibration and model exploration.

\subsection{Sensitivity Analysis}
\label{sec:SAresults}

\begin{figure*}
    \centering
    \includegraphics[width=\textwidth,trim={3cm 3cm 0 4cm},clip]{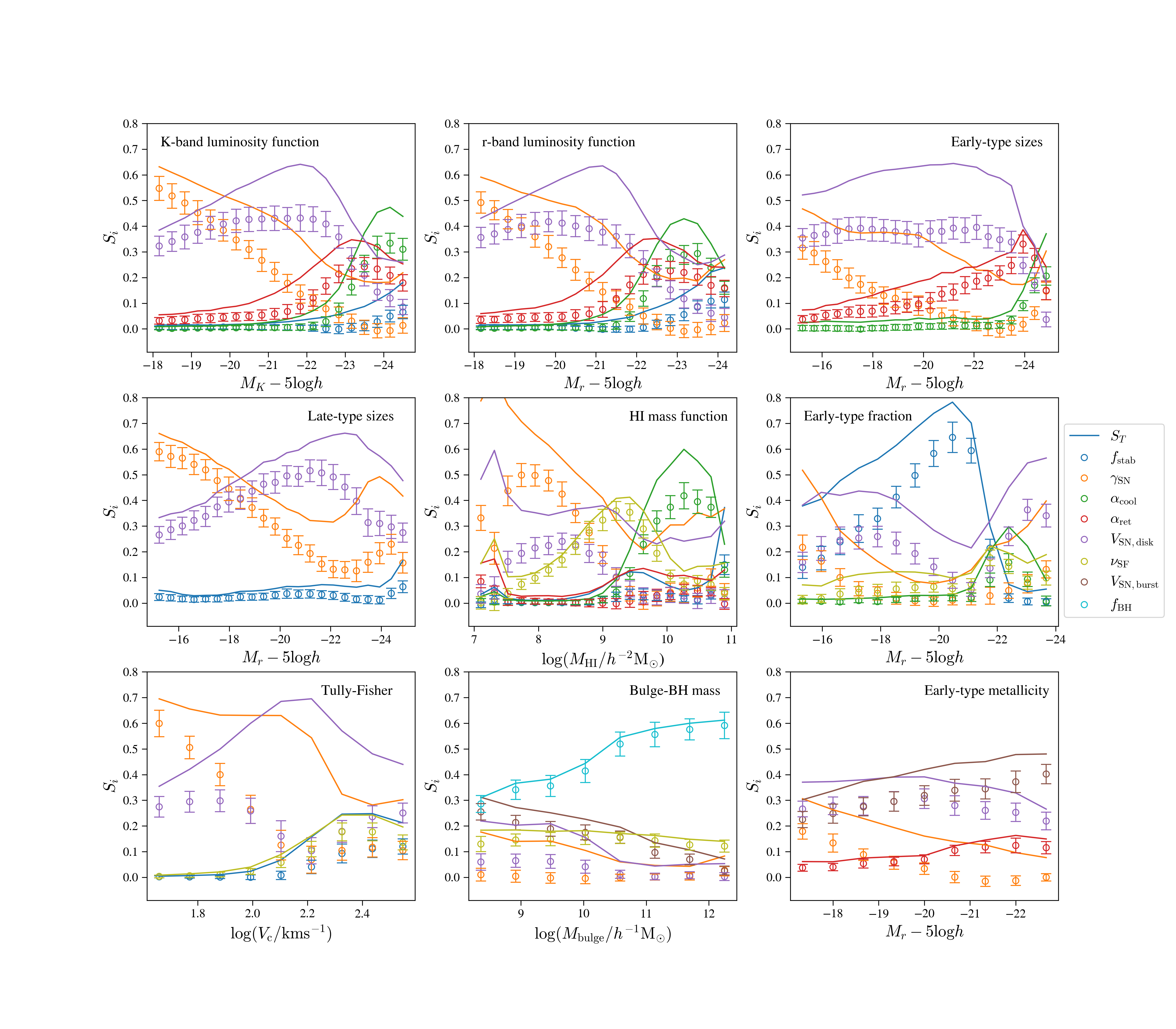}
    \caption{The emulator sensitivity to different parameters for each of the observables considered in this work; each panel shows a different observable, as labelled. Open circles indicate $S_{1}$ as described in the text, and solid lines represent $S_{T}$. For clarity, error estimates are shown for the $S_{1}$ calculation but not for $S_{T}$, although they are similar. Sensitivities for parameters which never exceed more than 10\% of the variance in any bin are not plotted.}
    \label{fig:EmulatorSA}
\end{figure*}

We apply the techniques described in \S~\ref{sec:SAdescription} to calculate the contribution of each parameter to the variance in each bin of the 9 constraints. The results are shown in Fig.~\ref{fig:EmulatorSA}. The open circles indicate the first order sensitivity index, $S_{1}$, which quantifies the proportion of the variance due to just one parameter. The total order sensitivity, $S_{T}$, is shown as solid lines, and indicates the proportion of the variance contributed by one parameter and its interactions with the other parameters. We can interpret the difference between the first order and total order sensitivity as a measure of the strength of the interaction between a given parameter and the other parameters. For clarity, we exclude parameters which never contribute more than 10\% of the variance to any bin. Both $f\textsubscript{burst}$ and $f\textsubscript{ellip}$ meet this condition, and so do not appear in the plots. 

We see that the dominant parameters for the majority of the model outputs are, perhaps unsurprisingly, the supernova feedback parameters. $V\textsubscript{SN, disk}$ and $\gamma\textsubscript{SN}$ account for the majority of the variance at the faint end of the K- and r- band LFs. At the bright end, $\alpha\textsubscript{cool}$, the parameter governing the strength of AGN feedback, contributes the largest proportion of the variance. The majority of the variance in the late- and early-type sizes, the Tully-Fisher relation, as well as the HI mass function is also contributed largely by the same two or three parameters. 

The early-type fraction is dominated by the threshold for disk instability, $f\textsubscript{stab}$, up until $M_{r}-5\text{log}h \approx -21$. At brighter magnitudes, disk instabilities become unimportant as mergers takes over as the main channel for building spheroidal components (see Husko et~al. 2021, in prep, for an exploration of the relative importance of different channels for the growth of galaxy stellar mass).  

The sensitivity analysis hence dispels one of the myths surrounding SAMs as it shows that the model cannot be made to fit to any arbitrary combination of datasets. To match the faint end of the K-band LF, we are strongly constrained in our choice of supernova feedback parameters, which contribute the vast majority of the variance to these bins. Our predictions of early- and late-type galaxy sizes, the HI mass function, the Tully-Fisher relation,  and the bright end early-type fraction are also then largely constrained, since the supernova feedback parameters dominate these outputs too. This is in line with the analysis performed by  \cite{Bower2010}, which reached similar conclusions.

The parameters also have clear physical interpretations, and are analogous to the parameters used in the subgrid physics models in hydrodynamic simulations \citep[e.g.][]{Crain2015, illustris2, illustris1}. The parametric model for supernova feedback can indeed be tuned to give a good match to the late-type galaxy sizes, but in doing so we are strongly constraining our fits to other datasets; the model does not include arbitrary parameters which allow for fine-tuning to an individual dataset without physical motivation or consequences for the fits to other datasets. 

\subsection{Calibration and dataset tensions}
\label{sec:calibration_tension_results}

We now apply the methods described in \S~\ref{sec:parameterFittingDescription} to calibrate the model to the datasets described in \S~\ref{sec:calibrationDatasetsDescription}, focusing on uncovering any tensions that exist between datasets. First, we aim to replicate a known tension in the model discussed in \cite{Bower2010} and \cite{Lacey2016}. This is the tension between reproducing late-type galaxy sizes and the galaxy LFs; these datasets have been found to prefer different values for the supernova feedback parameters. We can investigate this by adjusting the weightings applied to the residuals between our emulator prediction and each dataset (as in Eqn.~\ref{eq:MCMC}), and then performing an MCMC parameter search to see how the best-fitting parameter choices respond. 
    
\begin{figure*}
    \centering
    \includegraphics[width=\textwidth,trim={0cm 0.6cm 0 0cm},clip]{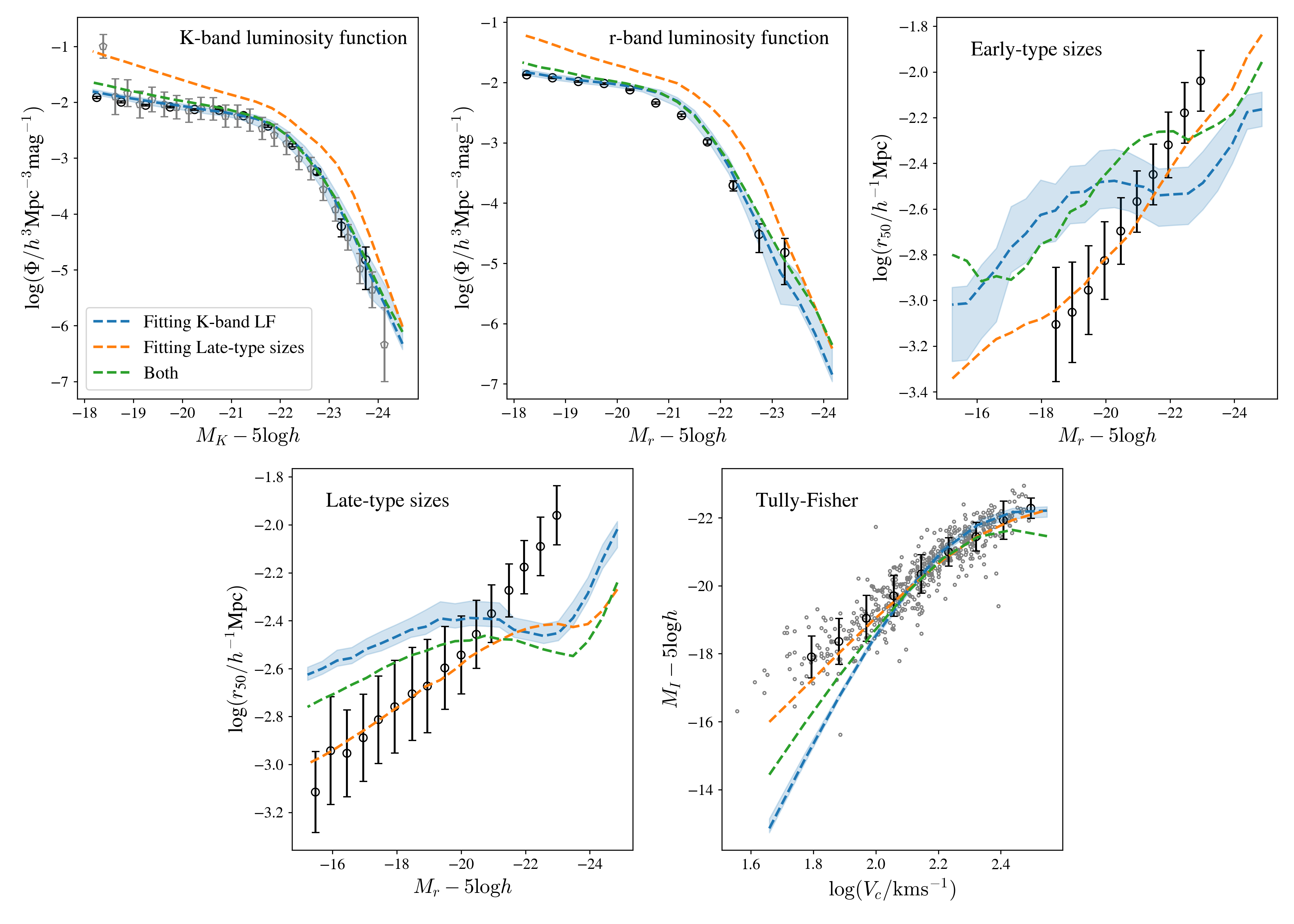}
    \caption{A comparison of the emulator predictions for fits to the K-band luminosity functions, the late-sizes, and a combination of the two (represented by different colour dashed lines). We fit to the data from \protect\cite{Driver2012} (\textit{black}) for the K-band LF, and \protect\cite{Shen2003} for the late-type sizes. The emulator predictions correspond to the best fit found from 20 MCMC chains, each 10,000 steps in length. The blue shading represents the 10-90th percentile errors when predicting a similar value in the holdout set. The black and grey datapoints represent the calibration data described in \S\ref{sec:calibrationDatasetsDescription}. For the K-band LF, we also compare to data from  \protect\cite{Kochanek2001} (\textit{grey}). For the r-band LF, we compare to data from \protect\cite{Driver2012}. For the early-type sizes we compare to data from \protect\cite{Shen2003}, and for the Tully-Fisher relation we compare to data from \protect\cite{deJong2000}.}
    \label{fig:TFLFcomparison}
\end{figure*}

In Fig.~\ref{fig:TFLFcomparison}, we show the emulator predictions for three sets of best-fitting parameters. In the first case, shown by the blue line, we weight only the residuals for the K-band LF. For the orange line, we weight only the size-luminosity relation for late-type galaxies, and the green line shows the results when weighting both datasets equally (i.e. both datasets have equal influence over the best-fitting parameter values). The shaded region is shown only around the fit to the K-band LF for clarity, and represents the 10-90th percentile error of the emulator when predicting similar values in the holdout set  (this gives a rough idea of the uncertainty of the emulator, but is certainly not an exact measure). We can clearly see the tension between these two datasets uncovered in an automatic and objective way; matching the sizes of faint late-type galaxies leads to an over-prediction of the LF at all luminosities by up to an order of magnitude. When matching both the K-band LF and the late-type galaxy sizes, we see an over-prediction in the faint-end of the LFs, and the sizes of faint late-types are over-predicted by a factor of $\sim 2$.  The early-type sizes and Tully-Fisher relation are also shown in Fig.~\ref{fig:TFLFcomparison}. Although no weighting was applied to these datasets in this exercise, we can see improved matches emerge naturally when we fit to the late-type galaxy sizes. We can gain some intuition for this behaviour from Fig.~\ref{fig:EmulatorSA}. As discussed, the Tully-Fisher relation, early- and late-type galaxies sizes, and the faint-end of the galaxy LF are highly sensitive to the choice of supernova feedback parameters,  $\gamma\textsubscript{SN}$ and $V\textsubscript{SN, disk}$ (which together account for $\sim90\%$ of the variance in the faint-end LFs and the sizes of faint late-type galaxies). Therefore we might expect that some tension would arise in trying to fit to a number of the above datasets at the same time.

\begin{figure*}
    \centering
    \includegraphics[width=\textwidth,trim={3cm 3.5cm 3.5cm 3.8cm},clip]{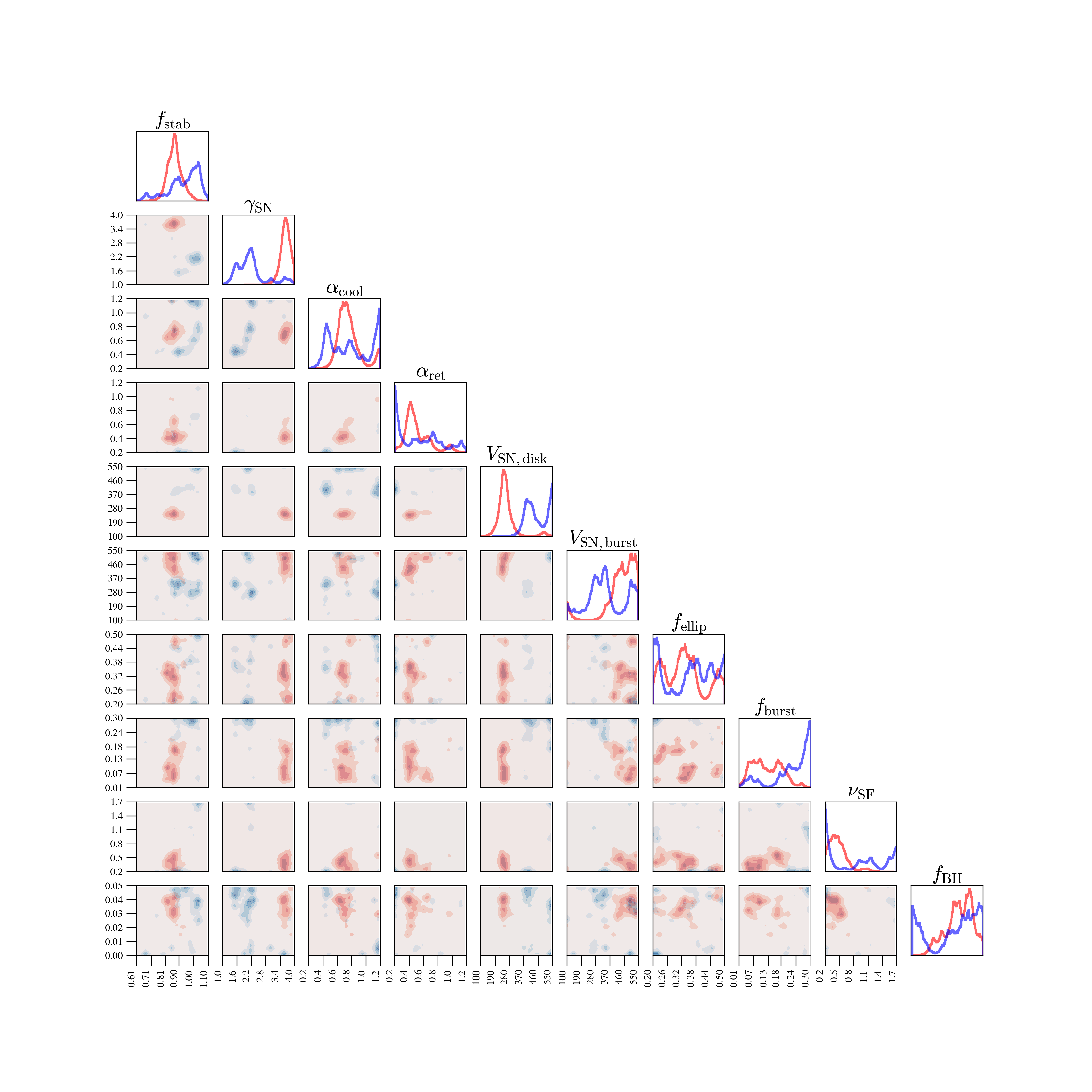}
    \caption{Accepted samples from 20 MCMC chains for fits to the K-band LF (\textit{red}), and both the K-band LF and the late-type galaxy sizes (\textit{blue}). The first 50\% of samples were discarded to allow for burn-in. The histograms show the marginalised distribution of the parameters. The ranges on each axis are the same as those quoted in Table \ref{tab:ParamRanges}. The shading gives a sense of the density, with darker colours corresponding to more densely sampled regions. The darkest regions correspond to the 25th percentile, and the lighter regions to the 50th and 75th percentiles.}
    \label{fig:TFLFcorner}
\end{figure*}

It is also informative to investigate how the acceptable regions of parameter space change as we introduce weightings to other datasets.  We demonstrate this for the tension between the LF/late-type sizes in Fig.~\ref{fig:TFLFcorner}. The shaded regions represent accepted samples from our 20 MCMC chains, each 10,000 steps in length, with the first 50\% of each chain
discarded to allow for burn-in. The red region corresponds to a fit to the K-band LF, and the blue region to  fits to both the K-band LF and late-type galaxy sizes. 
The shading gives a sense of the density of accepted samples i.e. the darker colours correspond to the more favoured parts of parameter space in this projection. The darkest regions correspond to the 25th percentile, and the lighter regions to the 50th and 75th percentiles. Also shown in Fig.~\ref{fig:TFLFcorner} are 1D histograms of the density of accepted samples. We find that, as in previous analyses, a reasonably large range of parameter 
values result in acceptable fits to a given constraint. This can be best understood \citep[as explained in][]{Bower2010} as the  
effect of the 
high dimensionality of the 
parameter space; 
though when plotted in projection down to 1 or 2 dimensions the space appears widely sampled, 
the higher dimensional acceptable region is reduced significantly. Also, some of the parameters produce degenerate effects (see for example Fig.~\ref{fig:OATVsnburst} in Appendix~\ref{appendix:ExtraFigures}, where we show the degenerate effects of the $f\textsubscript{stab}$ and $V\textsubscript{SN, burst}$ parameters). Nevertheless, we see that the K-band LF fit prefers somewhat higher values of $\gamma\textsubscript{SN}\approx3.6$ and lower values of $V\textsubscript{SN, disk}\approx200$kms\textsuperscript{-1}, in contrast to the fit to both the K-band LF and late-type sizes, where we find a preferred value of $\gamma\textsubscript{SN}\approx2.3$ and $V\textsubscript{SN, disk}$ at the top of the explored range at $\sim 550$kms\textsuperscript{-1}. Interestingly, there seems also to be a preference for lower values of $\nu\textsubscript{SF}$ to match the late-type galaxy sizes. We can understand this crudely by investigating the first-order effect associated with the $\nu\textsubscript{SF}$ parameter. Inspecting Fig.~\ref{fig:OATnusf} in Appendix~\ref{appendix:ExtraFigures}, we see that the $\nu\textsubscript{SF}$ parameter has a some effect on the bright-end of the K-band LF. This counteracts the enhancement from the higher value of $V\textsubscript{SN, disk}$, and also marginally improves the fit to the late-type galaxy sizes.

\begin{figure*}
    \centering
    \includegraphics[width=\textwidth,trim={0.5cm 0.5cm 0cm 0cm},clip]{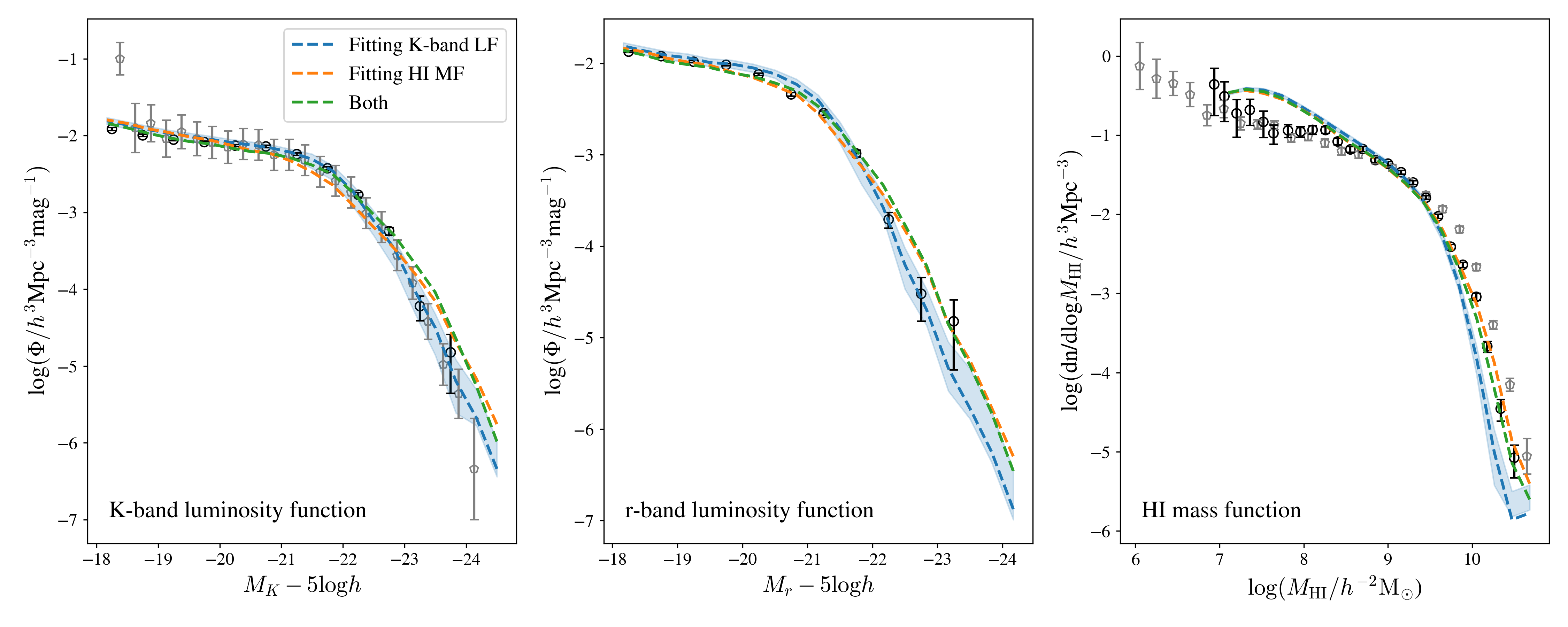}
    \caption{A comparison of the emulator predictions for fits to the K-band luminosity functions, the HI mass function, and a combination of the two (represented by the different colour dashed lines). The black and grey datapoints represent the calibration data described in \S\ref{sec:calibrationDatasetsDescription}. For the HI mass function, we fit to data from \protect\cite{Zwaan2005} (\textit{black}) and include data from \protect\cite{Martin2010} (\textit{grey}) for comparison.}
    \label{fig:LFHIcomparison}
\end{figure*}

Another tension arises between the HI mass function and the bright end of the K- and r-band LFs. This is shown in Fig.~\ref{fig:LFHIcomparison}. As before, the blue line corresponds to the fit to the K-band LF alone, the orange line to the fit using the HI mass function alone, and the green to a fit to both datasets. We can again propose (from our plot of the sensitivity indices, Fig.~\ref{fig:EmulatorSA}) that the main cause of this discrepancy is a tension in the choices for the AGN feedback parameter, $\alpha\textsubscript{cool}$, and the supernova feedback parameters. Indeed, when fitting the observational constraints individually, the fit to the K-band LF prefers a higher value for the AGN feedback parameter, with $\alpha\textsubscript{cool}\approx0.8$, whereas the fit to the HI mass function prefers $\alpha\textsubscript{cool}\approx0.5$.  We can also investigate how calibrating to both datasets shifts the parameter values. We do this as before with an MCMC exploration of the parameter space (see Fig.~\ref{fig:HItensionparams} in Appendix~\ref{appendix:ExtraFigures}). Fitting to both the K-band LF and the HI mass function (as compared with a fit just to the K-band LF) causes a shift in the preferred $V\textsubscript{SN, disk}$ to higher values. $\nu\textsubscript{SF}$, the parameter which controls the rate of quiescent star formation, shifts to the lowest values in the explored range, and the parameter $\alpha\textsubscript{ret}$, which is involved in gas return to halos following supernova feedback, becomes more strongly peaked, with the peak shifted to slightly higher values. 

\begin{figure}
    \includegraphics[width=\columnwidth,trim={0cm 0.65cm 0cm 0cm},clip]{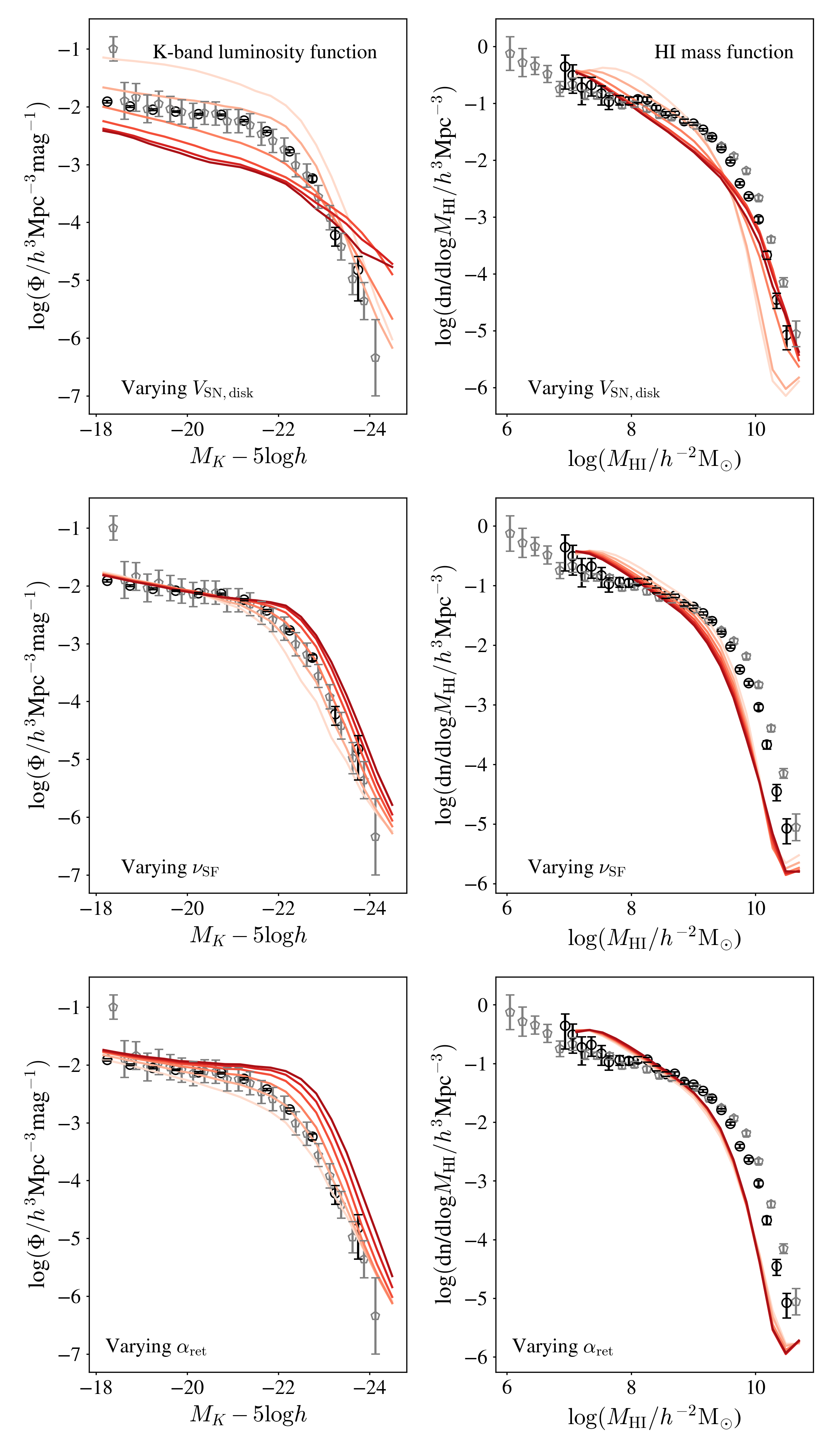}
    \caption{Emulator predictions for perturbing three key parameters around a fit to the K-band LF. The top row shows the result of varying the parameter $V\textsubscript{SN, disk}$ between 100 and 550 kms\textsuperscript{-1}, the middle row varies $\nu\textsubscript{SF}$ between 0.2 and 1.7 Gyr\textsuperscript{-1}, and the bottom row varies $\alpha\textsubscript{ret}$ between 0.2 and 1.2. Darker colours correspond to higher values of the varied parameter.}
    \label{fig:OATHILF}
\end{figure}

To understand this further, we investigate the first-order effects of the parameters ($V\textsubscript{SN, disk}$, $\nu\textsubscript{SF}$, and $\alpha\textsubscript{ret}$), perturbed around the fit to the K-band LF. We show the results in Fig.~\ref{fig:OATHILF}. We vary the parameters individually (`one-at-a-time') across their explored range, with lighter colors corresponding to lower parameter values. We can begin to understand the changes in the preferred parameter choices in terms of these transformations. When fitting both the HI mass function and the K-band LF, we find that there is a slight over-prediction of the bright-end of the LF. From these  one-at-a-time plots we can see that the increase in $V\textsubscript{SN, disk}$ causes an over-prediction at the bright-end of the LF, and a reduction in amplitude at the faint-end, but more accurately matches the high-mass end of the HI mass function. The HI mass function can be better matched at intermediate masses by a decrease in $\nu\textsubscript{SF}$. In  \texttt{GALFORM}, reducing $\nu\textsubscript{SF}$ has the effect of decreasing the rate of quiescent star formation in disks. As a result, lower values of this parameter provide a better fit to intermediate masses of the HI mass function, while simultaneously reducing the number density of the most luminous galaxies in the K-band LF, and so counteracting the enhancement due to the increase in the $V\textsubscript{SN,disk}$ parameter. We can further improve the match of the prediction for the LF to the observational data by increasing $\alpha\textsubscript{ret}$, which has little impact on the HI mass function but reverses some of the 'flattening' of the LF caused by the increase in $V\textsubscript{SN,disk}$. In previous galaxy formation models, using the WMAP-7 cosmological parameters, this tension has not been so apparent, but can also be seen between the b\textsubscript{J}-band LF and the HI mass function in \cite{Baugh2019}.

\begin{figure*}
    \centering
    \includegraphics[width=\textwidth,trim={0.5cm 0.5cm 0cm 0cm},clip]{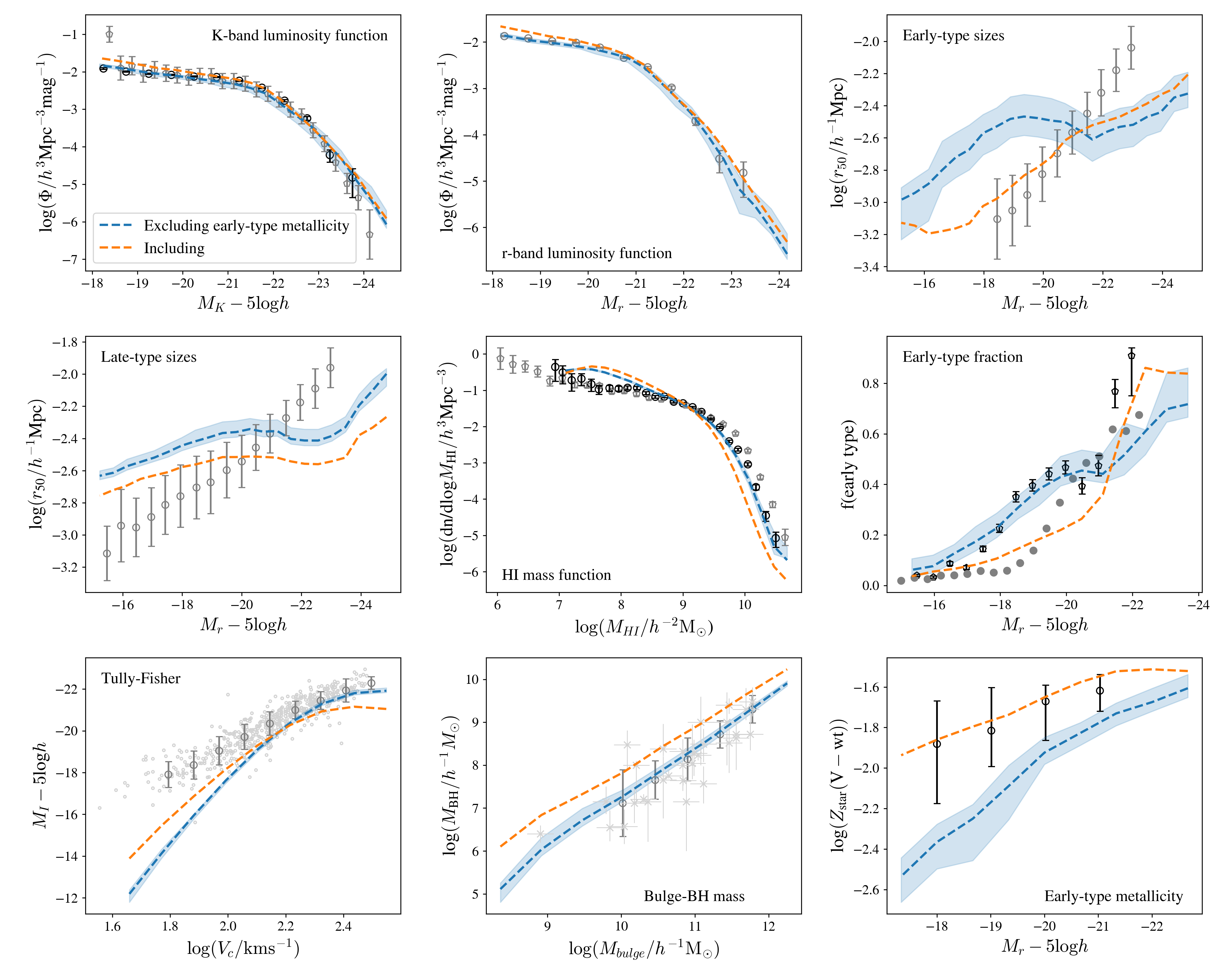}
    \caption{A comparison of the emulator predictions for fits to the K-band LF, the HI mass function, and the early-type fraction with and without including the early-type metallicity constraint (represented by different colour dashed lines, as labelled in the top left panel). The emulator predictions correspond to the best fit found from 20 MCMC chains, each 10,000 steps in length. In both cases, all included constraints were equally weighted. The data described in  \S\ref{sec:calibrationDatasetsDescription} is shown in black and grey. For the Bulge-BH mass relation we compare to data from \protect\cite{Haring2004}, for the early-type fraction we fit to data from \protect\cite{Moffett2016} and compare to data from \protect\cite{Gonzalez2009}, and for the early-type metallicity we compare to data from \protect\cite{Smith2009}. Black data points indicate that the data was used for fitting, grey data points are included for comparison. }
    \label{fig:ETMcomparison}
\end{figure*}

\begin{figure}
    \centering
    \includegraphics[width=\columnwidth,trim={0.8cm 0.75cm 1.2cm 1.3cm},clip]{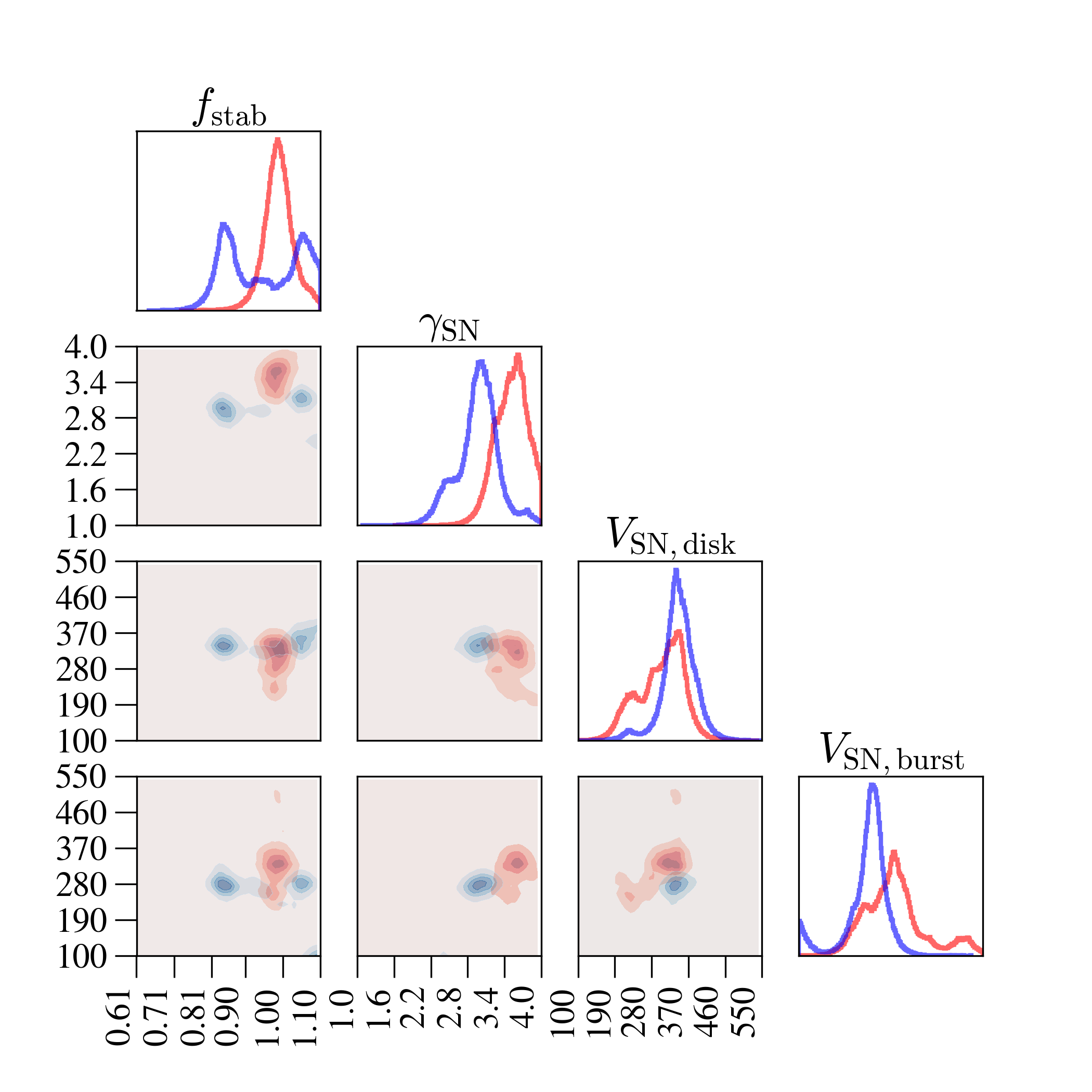}
    \caption{Accepted samples from 20 MCMC chains for fits to the K-band, the HI mass function, and the early-type fraction with (\textit{blue}) and without (\textit{red}) including the early-type metallicity constraint for a few key parameters. The shading gives a sense of the density of the samples, and the histograms show the distribution of each parameter in 1D projection. The darkest regions correspond to the 25th percentile, and the lighter regions to the 50th and 75th percentiles.}
    \label{fig:EMparams}
\end{figure}

Our approach also allows us to uncover a significant tension between the bright end of the LFs, the early-type fraction, the HI mass function, and the early-type metallicity. We demonstrate this in Fig.~\ref{fig:ETMcomparison}, where we compare a fit found by calibrating to the K-band LF, HI mass function, and the early-type fraction with and without including the early-type metallicity constraint (note that we do not fit to datasets shown in grey). Including the early-type metallicity has a significant effect on the best-fitting parameter values; it generally improves the fits to the galaxy sizes, and degrades the fit to the early-type fraction (at least when considering the \cite{Moffett2016} data) and the HI mass function. We investigate the impact on the acceptable region of parameter space in Fig.~\ref{fig:EMparams}, where we show the key changes induced by including the early-type metallicity constraint. The red region shows the fit to the K-band LF, HI mass function, and early-type fraction, and the blue region also includes the early-type metallicity. We find that there is a reconfiguration of the supernova feedback parameters, $\gamma\textsubscript{SN}$, and $V\textsubscript{SN, burst}$ to match the early-type metallicity. This reconfiguration provides better fits to the galaxy sizes, while degrading the fit to the HI mass function, which is also very sensitive to the choice of $\gamma\textsubscript{SN}$. The fits found when we choose not to include the early-type metallicity constraint are very similar to those found in \cite{Lacey2016, Baugh2019}, with over-predictions for the sizes of faint early-type galaxies, good fits to the HI mass function, and an under-prediction of the metallicity of faint early-type galaxies. Including the early-type metallicity constraint,  however, moves us to a different region of parameter space for this updated version of the \texttt{GALFORM} code.

Another key shift is in the preferred value of $f\textsubscript{stab}$; the preference for lower values of $f\textsubscript{stab}$ leads to a suppression of the early-type fraction at intermediate luminosities. At these luminosities, disk instabilities are the main channel for building up spheroid components and decreasing $f\textsubscript{stab}$ limits the number of disk instabilities (see Husko et~al., in prep). Although $f\textsubscript{stab}$ does not appear in the early-type metallicity sensitivity analysis (as shown in Fig.~\ref{fig:EmulatorSA}), this is because the sensitivity indices are dominated by the strong effects of the supernova feedback parameters. A lower $f\textsubscript{stab}$ does increase the early-type metallicity but to a far lesser extent than the supernova feedback parameters, and so gives a more exact match to the observational data. 

In our analysis so far, we are perhaps making the mistake of attempting to understand a non-linear model in terms of just first order, one-at-a-time changes to the parameters. Indeed, this is one of the key weaknesses of traditional `chi-by-eye' parameter fitting. However, as shown in Fig.~\ref{fig:EmulatorSA}, we can justify this mode of investigation; the majority of the variance due to a given parameter is generally due to just its first-order effect. $\nu\textsubscript{SF}$, $\alpha\textsubscript{ret}$, $\alpha\textsubscript{cool}$ and $f\textsubscript{SMBH}$ only have weak higher order variance contributions. In the cases where this assumption is less valid, for example in the case of the parameter $\gamma\textsubscript{SN}$ and $V\textsubscript{SN,disk}$, this can be understood straightforwardly with reference to Eqn.~\ref{eq:SNfeedback}; these parameters directly interact in the implementation of supernova feedback. It is striking how much of the variance is due to the parameters' first order effects. The outlier is $f\textsubscript{stab}$, which has strong higher-order interactions and is not directly coupled to the other parameters in any equation.

\subsubsection{Best-fitting model}
\label{sec:bestfit}

We can now re-calibrate the \texttt{GALFORM} model across all  constraints to produce an estimate of the best-fitting parameters. As we have seen, there is no single choice of parameters which can reproduce all of the constraints, and we have to decide during the calibration which datasets we would like to give more or less weighting. The ideal of automatically calibrating a semi-analytic model is therefore a difficult one to realise; we will always have to make trade-offs in how we fit to the various datasets. As described in \S~\ref{sec:calibrationDatasetsDescription}, we can do this in a semi-automatic way using the heuristic weighting scheme. 

We have seen that there are a number of trade-offs or tensions to consider when aiming to find a best-fitting model. Fitting to the late-type galaxy sizes, the Tully-Fisher relation, or the HI mass function generally degrades the fit to the K- and r-band LFs. We have also seen that trying to reproduce the early-type metallicities worsens the fit to the \cite{Moffett2016} data for the early-type fraction, and worsens the fit to the high-mass end of the HI mass function. On the other hand, other observational constraints are more easily fitted; since the bulge-BH mass relation is largely dependent solely on the $f\textsubscript{SMBH}$ parameter, and this has very little influence on other observables, fitting this constraint is trivial.

With these considerations in mind, we choose heuristic weights such that the r- and K-band LFs are strongly weighted. We know from our previous analysis that there will be trade-offs between both the bright- and faint-ends of the luminosity functions, but we require good fits to both. Therefore we doubly weight both of these constraints when calculating the MAE given in Eqn.~\ref{eq:MCMC} (i.e. by setting $W_{i} = 2$ for each observable). Since the late-types sizes, early-types sizes, and the Tully-Fisher relation are important constraints, but lead to compromised LF fits, we apply single weighting to all these constraints (i.e. $W_{i} = 1$). We also give a single weighting to the early-type metallicity since this trades-off against the bright end of the luminosity function and the high mass end of the HI mass function. Since the HI mass function is an important constraint, but as we are aware that it generally degrades the fit to the bright end of the luminosity function, we give this constraint a triple weighting. This is to ensure that more total weight is applied to the K- and r-band LFs in combination. We apply a single weighting to the early-type fraction; we have seen that this fit is in strong tension with the early-type metallicities and sizes. 

We run 100 MCMC chains with our emulator, each 10,000 steps in length. We find that the minimum MAEs (as computed using the emulator) obtained with each chain lie in the range $\sim0.15-0.20$; since this range is similar to the out-of-sample accuracy of the emulator, and so in principle we cannot discern which parameter sets give the best fit to the observational data with the emulator alone, we evaluate these 100 minimum MAE parameter sets with the \texttt{GALFORM} code.
\begin{table}
	\centering
	\caption{The best-fitting parameters (as measured by MAE, Eqn.~\ref{eq:MCMC}) found by using MCMC combined with our emulator. For reference the last column gives the parameter values used by \protect\cite{Baugh2019}. The first column indicates the set of parameters with the lowest MAE, and the second column indicates the parameter ranges of the 50 best runs of the 100 MCMC chains, again selected by MAE as described in the text. 
	}
	\label{tab:BestFiticool6Params}
	\begin{tabular}{lccc} 
	\hline
	Parameter & This work & Range & Baugh19  \\
	\hline
	 $f\textsubscript{stab}$ & 0.79 & $0.73 - 1.00$ & 0.90 \\
     $\alpha\textsubscript{cool}$& 0.84 & $0.66 - 1.16$ & $0.80$ \\
     $\alpha\textsubscript{ret}$& 0.59 & $0.32 - 0.86$ & $1.00$ \\
     $\gamma\textsubscript{SN}$& 2.24 & $2.05 - 2.72$ & 3.40 \\
     $V\textsubscript{SN, disk}\, [\mathrm{kms}^{-1}]$ & $489$ & $368 - 541$ & 320\\
     $V\textsubscript{SN, burst}\, [\mathrm{kms}^{-1}]$ & $284$ & $230 - 292$ & 320\\
     $f\textsubscript{burst}$& 0.25 & $0.12 - 0.30$ & $0.05$ \\
     $f\textsubscript{ellip}$& 0.20 & $0.20 - 0.39$ & $0.30$ \\
     $\nu\textsubscript{SF}$ $[\mathrm{Gyr}^{-1}]$ & $0.20$ & $0.20 - 0.33$ & $0.74$ \\
     $f\textsubscript{SMBH}$& 0.003 & $0.001 - 0.004$ & $0.005$ \\
	    		\hline
	\end{tabular}
\end{table}

\begin{figure*}
    \centering
    \includegraphics[width=\textwidth,trim={0.5cm 0.5cm 0cm 0cm},clip]{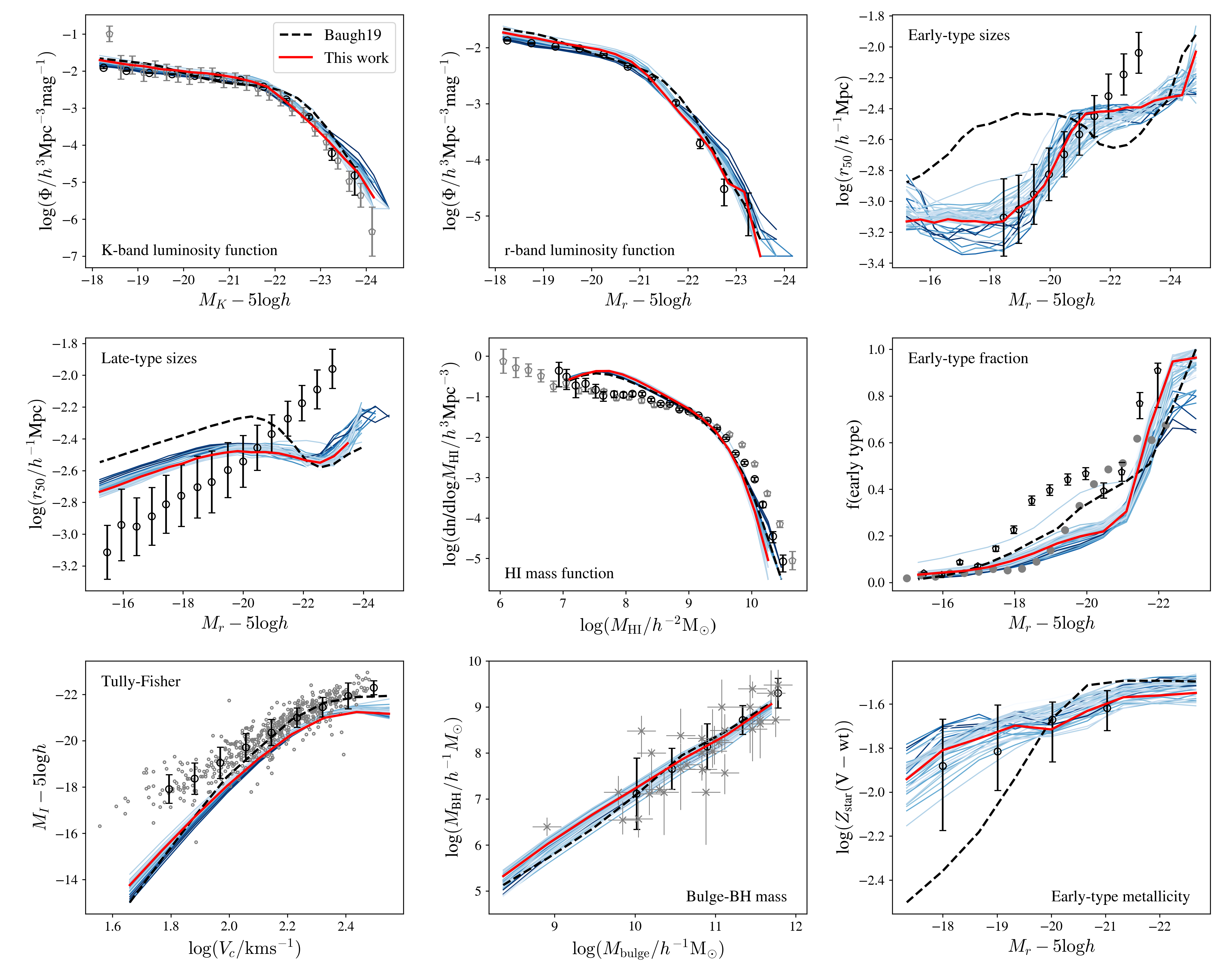}
    \caption{The \texttt{GALFORM} evaluations of the best-fitting parameters found with 100 MCMC chains, each 10,000 samples in length, using the constraint weightings described in the text. Here we plot a sample of the best 50 runs, as measured by MAE. The red line indicates the parameter set with the lowest MAE. The remaining 49 runs are plotted in blue, with darker shades indicating small residuals to the HI mass function. Note therefore that runs with the smallest residuals to the HI mass function over-predict the bright-end of the K- and r-band LFs. The black dashed line shows the Baugh19 model. The data described in \S\ref{sec:calibrationDatasetsDescription} is shown in black and grey. We calibrate to the data shown in black.
    }
    \label{fig:BestFiticool6}
\end{figure*}

The best-fits  are shown in Fig.~\ref{fig:BestFiticool6}. Here we plot the best 50 sets of parameters from the 100 MCMC chains, as evaluated with the \texttt{GALFORM} code. These runs have very similar MAEs, covering the range $0.16-0.18$, while the runs not shown cover the range $0.18-0.22$, which is slightly wider than the range predicted by the emulator, but within the expected emulator error (0.04 in this weighting scheme). The solid red line indicates the run with the lowest MAE, and the blue lines show the remaining 49 runs. The shading on these lines indicates the size of the residuals between the model and the HI mass function, with darker lines indicating smaller residuals, and demonstrates that the parameter choices which provide the best fits to the HI mass function over-predict the bright-end of the LFs. The black dashed line shows the statistical galaxy properties of the model presented in \cite{Baugh2019} (hereafter Baugh19). In Table~\ref{tab:BestFiticool6Params} we show the set of parameters with the lowest MAE to the observational data (corresponding to the red line in Fig~\ref{fig:BestFiticool6}), the parameter range of the best 50 parameter sets, and compare with the parameters adopted in Baugh19 for an older version of the model. We reiterate, however, that the best-fit parameters are just one realization out of many possible choices due to the 
degeneracies between the parameters, and the effect of calibrating to multiple datasets. Also, the ranges shown in Table~\ref{tab:BestFiticool6Params} do not indicate that any choice of parameters within these ranges will yield an equivalent fit; the value of one parameter will constrain the choices for the other parameters, hence the reason for giving an example of a best-fitting set of parameters. We find that some parameters, such as $\nu\textsubscript{SF}$ and $\gamma\textsubscript{SN}$ are constrained to a tight range of values, whereas others, such as $f\textsubscript{stab}$ can be drawn from a large fraction of the explored range.

Calculating the mean absolute error of the best-fitting model, and the Baugh19 model, using the same procedure as described in \S\ref{sec:parameterFittingDescription} (and recalling that we scale each output so that the data lie in the range [0,1]), we find that at least under this metric the new model is a better fit to the data. Over all the datasets, the new best-fit found in this work gives an MAE of 0.16 vs. an MAE of 0.20 for the Baugh19 model. We note that the MAE for the model used in Baugh19 is within the range of the minimum MAE reached by the 100 MCMC chains. The reduced MAE of the new best fitting model compared to the Baugh19 model is mainly due to large improvements in the fits to the early-type galaxy sizes and their metallicities, while the fits of the new model to the early-type fraction and Tully-Fisher relation are slightly worse.

As shown in Fig~\ref{fig:BestFiticool6}, we find that our model provides a slightly better fit to the K- and r-band LFs than the Baugh19 model\footnote{Baugh et~al. concentrated on reproducing the $b_{\rm J}$-band luminosity function, and the HI mass function, and did not consider the $r$-band LF.}. For the updated model presented in this work, we find an MAE of 0.05 vs. 0.08 for the Baugh19 model in the K-band and 0.04 vs. 0.06 in the r-band. The galaxy sizes are an improvement over previous iterations of the \texttt{GALFORM} model, particularly the early-types, which are now more qualitatively similar to the observational data in that they are monotonically increasing with luminosity (at least in the range of the data), whereas the Baugh19 model features a marked dip at intermediate magnitudes and significant over-prediction at fainter magnitudes, differing from the observed sizes by a factor of $\sim3$. The MAEs in this case are also significantly lower for the new model: for the late-type galaxies we find an MAE of 0.14 in this work vs. 0.21 for the Baugh19 model, and 0.09 vs. 0.39 for the early-type sizes. This difference is largely due to the different choices for the $\gamma\textsubscript{SN}$ parameter. Here, we find a preference for much lower values of $\gamma\textsubscript{SN}$, in the range $2.05-2.72$, vs. 3.40 for the Baugh19 model. Reducing this parameter significantly weakens the effect of supernova feedback in low-mass galaxies, leading to smaller sizes \citep[see figure C10 of][]{Lacey2016}. Interestingly, the preferred  $\gamma\textsubscript{SN}$ we recover is much closer to the value expected from energy conservation arguments, $\gamma\textsubscript{SN} = 2$ \citep{larson1974, Lagos2013}.

The fit to the HI mass function is slightly worse than the fit found in the Baugh19 version of the model (with MAEs of 0.09 vs 0.08); a better fit would come at the expense of a more severe over-prediction of the bright-end of the luminosity function as previously discussed, and as shown by the shading of the blue lines in Fig.~\ref{fig:BestFiticool6}. As we have seen in Fig.~\ref{fig:ETMcomparison}, we are able to produce better matches to the HI mass function and the luminosity functions if we exclude the early-type metallicity and galaxy sizes constraints (the fits found in this case are much more similar to the Baugh19 model, with similarly high $\gamma\textsubscript{SN}$ in the range $\sim 3.2 - 3.8$, as shown in Fig.~\ref{fig:EMparams}). Our fit to the early-type metallicities is an improvement over the prediction of the Baugh19 version of the model, where the MAE of our model is 0.15 vs. 0.55 for the Baugh19 model. However, our early-type metallicities fit comes at the cost of slightly degrading the fit to the early-type fraction (0.13 vs. 0.10). Our fit to the Tully-Fisher relation is worse than in the Baugh19 model, with an MAE of 0.28 vs. 0.17, though we have demonstrated that we can retrieve a fit more similar to Baugh19 by giving less weight to the early-type metallicity constraint (again as shown in Fig.~\ref{fig:ETMcomparison}).

\subsubsection{Predictions for cosmic star formation history}

\begin{figure}
    \includegraphics[width=\columnwidth,trim={0cm 0.65cm 0cm 0.2cm},clip]{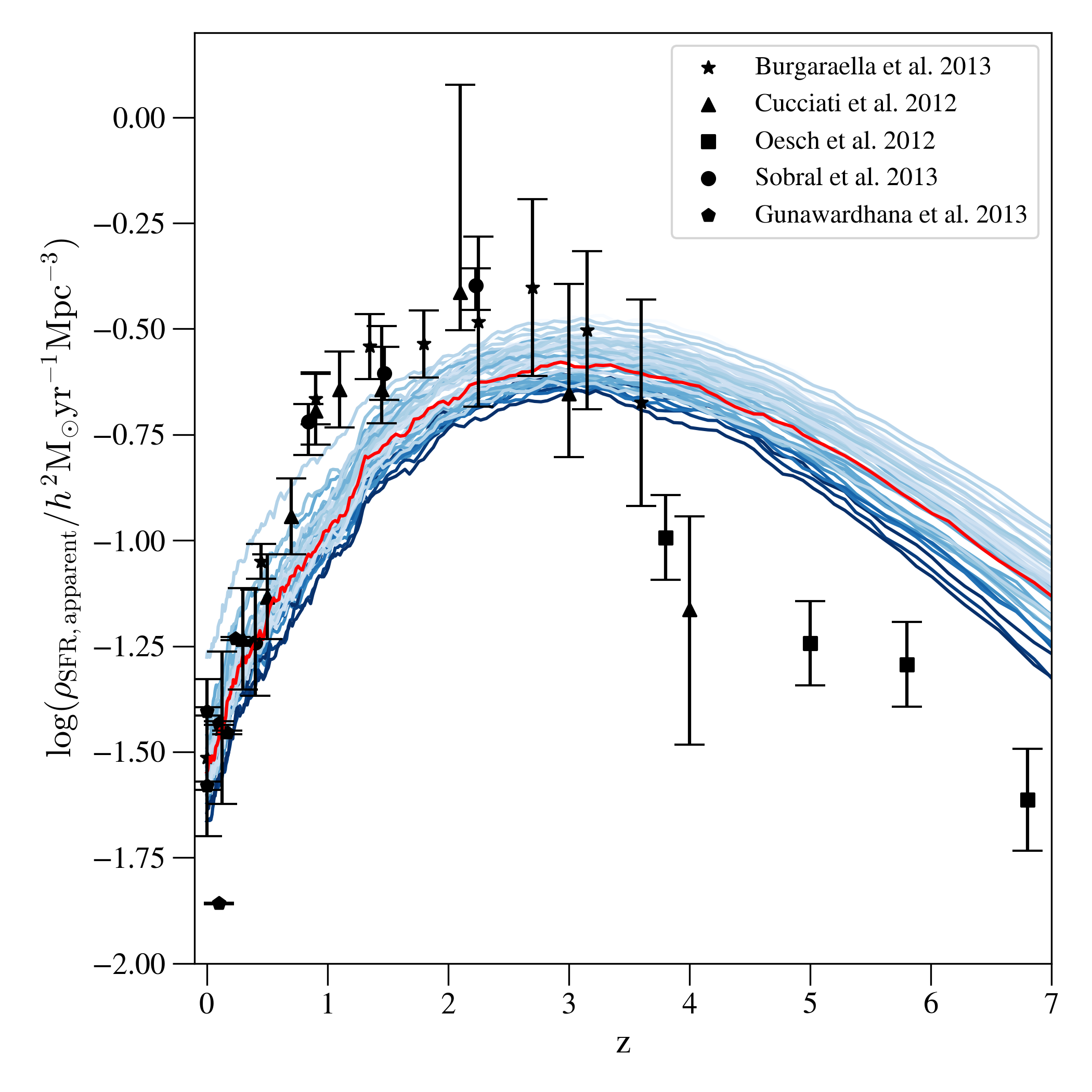}
    \caption{The apparent SFRD predictions for the \texttt{GALFORM} model evaluations shown in Fig.~\ref{fig:BestFiticool6}. The red line indicates the predictions for the best-fit parameters (as calculated by MAE), while the blue lines indicate the remaining 49 runs. These lines are shaded according to the model's residuals to the HI mass function, with darker shades indicating smaller residuals. We compare to observational data from \protect\cite{Burgarella2013, Cucciati2012, Oesch2013, Sobral2013, Gunawardhana2013}. Note that these data were not used in the fitting. 
    A correction has been applied to the predicted SFRD in bursts to give an apparent SFRD, as described in the text.}
    \label{fig:SFRD}
\end{figure}

We have calibrated  \texttt{GALFORM} to low-redshift constraints and now investigate the  predictions for the evolution of the star formation rate density (SFRD) with redshift. To do this, we evaluate the SFRD with redshift for the sets of parameters corresponding to the \texttt{GALFORM} runs shown in Fig.~\ref{fig:BestFiticool6}. Fig.~\ref{fig:SFRD} shows the SFRD predictions for these parameter choices. Since \texttt{GALFORM}  assumes a mildly top-heavy initial mass function (IMF) for stars formed in starbursts, we apply an approximate correction to give the SFR which would be inferred assuming a Kennicutt IMF \citep{Kennicutt1983} by weighting the starburst SFR by a factor of 1.9 \citep[as in][]{Lacey2016}. The curves therefore represent an apparent SFRD which can be compared with observational estimates which assume a solar neighbourhood IMF. Interestingly, we see that the spread of the model predictions only increases slightly as we move out to larger redshifts. This suggests that the low-redshift calibration datasets actually constrain the redshift evolution of the model reasonably well.

\section{Discussion}
\label{sec:Discussion}

We have presented a method for efficiently calibrating and exploring a SAM of galaxy formation across a wide range of outputs. In doing so, we have uncovered a number of tensions between datasets: for example, in Fig.~\ref{fig:ETMcomparison}, we found that on relaxing the requirement for a good fit to the early-type metallicities, we recovered a fit very similar to those found in \cite{Baugh2019} and \cite{Lacey2016}. By increasing the weight given to the early-type metallicity constraint, we moved to a new region of parameter space, changing our fit to the early-type fraction and early-type sizes. Tensions such as this point to either deficiencies in the model, or a discrepancy between the observational datasets. For example, again in Fig.~\ref{fig:ETMcomparison}, we see that the early-type fraction fit to the \cite{Moffett2016} data (shown in black) degrades when we include the early-type metallicity constraint. However, in this case the fit is then in better agreement with the \cite{Gonzalez2009} data (shown in grey). Similarly, for the HI mass function, the \cite{Zwaan2005} and \cite{Martin2010} datasets do not agree with one another, differing by up to a factor of five in abundance at high masses. 

In other cases, we can see a clearer deficiency in the \texttt{GALFORM}  predictions. For example, in Fig.~\ref{fig:TFLFcomparison} we show the effect of fitting to the K-band LF or the late-type galaxy sizes, or both together. We see that even when we fit only to the late-type sizes constraint, we are still not able to recover the observed monotonic increase in radius with increasing luminosity. Clearly, this suggests that the  treatment of the galaxy disk-sizes in \texttt{GALFORM} needs to be improved.

The emulation method presented here contrasts with previous work; most  emulators have focused on reducing the parameter space by using more approximate emulators, but with robust uncertainty measures, to iteratively reduce the volume of parameter space which could plausibly produce good fits to the data. \cite{vandervelden2021}, for example, used a total of 3000 runs over three waves to calibrate the \texttt{MERAXES} SAM to the stellar mass function. We have focused instead on maximizing the accuracy of our emulator of \texttt{GALFORM} across the whole parameter space. Our aim is to build an emulator which allows us to explore a wide range of calibration datasets, and different combinations of these datasets. As shown in Fig.~\ref{fig:EmulatorPerformance}, our emulator performs well: most of the key constraints are tightly predicted. 

In this vein, we have discounted the observational error bars to facilitate model exploration. In \S~\ref{sec:bestfit} we calibrated our model to the full set of observational datasets under consideration. However, since we did not include observational errors and used an absolute error metric, it is difficult to give meaningful error bars around our estimates of the best-fitting parameters quoted in Table~\ref{tab:BestFiticool6Params}. As previously mentioned, SAM calibration involves making trade-offs between certain observational constraints; often the best-fitting model is calibrated in a way which is poorly defined. We have attempted to reproduce and elucidate this process in an automatic way through a heuristic weighting scheme. We aim to investigate a more robust calibration analysis in the future with an improved treatment of the observational errors. 

Similarly, our approach could be extended to include a more robust measure of the emulator's uncertainty in reproducing  \texttt{GALFORM} outputs. When emulating a set of model outputs, we should ideally account for \textit{epistemic} and \textit{aleatoric} uncertainty. \textit{Epistemic} uncertainty refers to the uncertainty associated with the emulator's parameters (in this case, the weights of the neural network), and \textit{aleatoric} uncertainty refers to uncertainty inherent in the data generating process (for example, the sampling noise on the \texttt{GALFORM} outputs). Our approach does not currently model the epistemic uncertainty on the emulator's weights, but instead acts to reduce it by averaging over a number of individual estimates provided by the neural networks in our ensemble. It is possible therefore that we are discarding regions of the parameter space which potentially contain reasonable fits to the data. However, we are somewhat protected from this scenario in that the regions which are most difficult for our emulator to model are regions which produce `unusual' or `undesirable' outputs (e.g. such as LFs without a clear exponential break), which are unlikely to be good matches to the observations anyway. Nevertheless, ideally we would like our emulator to return an estimate of its uncertainty (both the uncertainty in the emulator's weights and uncertainty inherent in the data-generating process).  \texttt{GALFORM} is a deterministic code, but we are still limited by the noise associated with sampling from a relatively small population of galaxies at high masses. Bayesian neural networks \citep{Neal1994, Bishop1997} are a class of models which seek to incorporate epistemic and aleatoric uncertainty into the deep learning framework; these networks often apply independent Gaussian prior distributions over model weights, and model the outputs themselves as distributions. We believe this may be a promising line of inquiry to combine the power of the neural network's adaptive basis functions with the uncertainty quantification of a full Bayesian analysis. 

Another appealing method is the deep kernel learning approach \citep{Wilson2016}. Here, a deep neural network is employed to transform the inputs to the kernel of a Gaussian process regression, and it has been shown to outperform both the plain Gaussian process model and the plain deep neural network in a number of cases \citep[e.g.][]{Wilson2016, patacchiola2020} while also providing robust uncertainty estimates. Here, the deep neural network can be thought of as a feature extractor which reduces the number of features input into the Gaussian process kernel and so allowing it to better generalize to higher dimensional inputs.

In Fig.~\ref{fig:EmulatorPerformanceSamples}, we demonstrated that we could improve the performance of our emulator as much as 10\% by averaging over 10 neural networks, rather than using just one. It may be interesting to investigate this avenue further. Our method used a simple average, but if a selection of machine learning algorithms are able to give errors which are not strongly correlated (i.e. some fit better to certain examples than others), it may be possible to use a more sophisticated approach to incorporate the respective advantages of a number of different algorithms \citep[see e.g.][]{Maclin2016}.

 We have proposed a number of ways to investigate the \texttt{GALFORM} model with our emulator. We can use sensitivity analysis techniques to evaluate the effect of different parameters, and since the emulator is extremely fast, we can manually explore the outputs in detail. It may also be possible to use symbolic regression such as the proprietary software \texttt{EUREKA} \citep[as described in][]{Dubcakova2011} or sparse regression-based methods (e.g. \cite{Rudy2019}) to generate closed-form expressions of the neural network outputs if desired (i.e. an estimate of the functional form of the outputs). \cite{Cranmer2019}, for example, applied symbolic regression techniques in conjunction with graph neural networks to extract equations from cosmological simulations.

\section{Conclusions}
\label{sec:conclusion}

We have implemented a deep learning approach to emulate the \texttt{GALFORM} SAM. We trained an ensemble of deep learning algorithms to approximate the full model  using just 930  evaluations of \texttt{GALFORM}. We used this to explore the parameter space of \texttt{GALFORM}, and to calibrate the model parameters to a wide array of observations. Typically the exploration of a model parameter space and the determination of a best-fitting set of parameter requires many more than 930 explicit full calculations. Our emulator is remarkably accurate, particularly in regions of the parameter space for which the model gives outputs which are close to matching the observed Universe.  

We used sensitivity analysis to quantify the influence of different parameters on the model outputs, to better understand which parameters are of greatest importance in fitting to different observations (see \citealt{Oleskiewicz2019}). Here, as shown in Fig.~\ref{fig:EmulatorSA}, we found that the majority of the variance is due to just a few key parameters, which leads to tension when trying to calibrate to multiple observational datasets.

We explored the tensions between the use of different observational datasets further, using MCMC to fit the emulator output to observational data with a heuristic weighting scheme. This allowed us to reproduce the known  tension between the faint-end galaxy LFs in the K- and r-bands and the late-type galaxy sizes, and to uncover a number of others. Furthermore, we used the same technique to find a global fit to the observational datasets, finding a set of parameters which provide an improved fit to the early-type galaxy sizes and metallicities as compared with an earlier version of the \texttt{GALFORM} code presented in \cite{Baugh2019}.

We intend to apply our emulation approach to calibrate  \texttt{GALFORM} using the observed galaxy redshift distribution to generate mock galaxy catalogues for the DESI bright galaxy survey \citep{desi}.  This requires model outputs over a large number of redshifts, which makes running \texttt{GALFORM} more  computationally expensive. We are motivated therefore to reduce the required number of model evaluations as much as possible; calibrating the model across this redshift range would be prohibitively expensive for direct MCMC methods, and very difficult to achieve by-eye. Our emulator is ideally suited to this task; we have demonstrated that we require very few runs to achieve good accuracy, and that we are able to emulate over a wide range of outputs.

We believe our approach to be an inexpensive, intuitive and accurate alternative to other emulation techniques in the literature, and that this method will serve as an invaluable tool in quickly exploring and calibrating SAMs, and for the rapid assessment of the implications of changes to the underlying model. 

\section*{Acknowledgements}
EJE  is supported by a PhD Studentship from the Durham Centre for Doctoral Training in Data Intensive Science, funded by the UK Science and Technology Facilities Council (STFC, ST/P006744/1) and Durham University. CMB and CGL acknowledge the support from STFC  (ST/T000244/1). This work used the DiRAC@Durham facility managed by the Institute for Computational Cosmology on behalf of the STFC DiRAC HPC Facility (\hyperlink{www.dirac.ac.uk}{www.dirac.ac.uk}). The equipment was funded by BEIS capital funding via STFC capital grants ST/K00042X/1, ST/P002293/1, ST/R002371/1 and ST/S002502/1, Durham University and STFC operations grant ST/R000832/1. DiRAC is part of the National e-Infrastructure.

\section*{Data Availability}

The observational datasets, \texttt{GALFORM} outputs, parameter values, and best-fit parameter values underlying this article will be shared on reasonable request to the corresponding author.




\bibliographystyle{mnras}
\bibliography{library} 



\appendix
\section{Supplementary figures}
\label{appendix:ExtraFigures}

Here, we provide some additional figures figures to provide further illustration of points  discussed in the main text.

\begin{figure}
    \centering
    \includegraphics[width=\columnwidth,trim={0.5cm 0.5cm 0cm 0cm},clip]{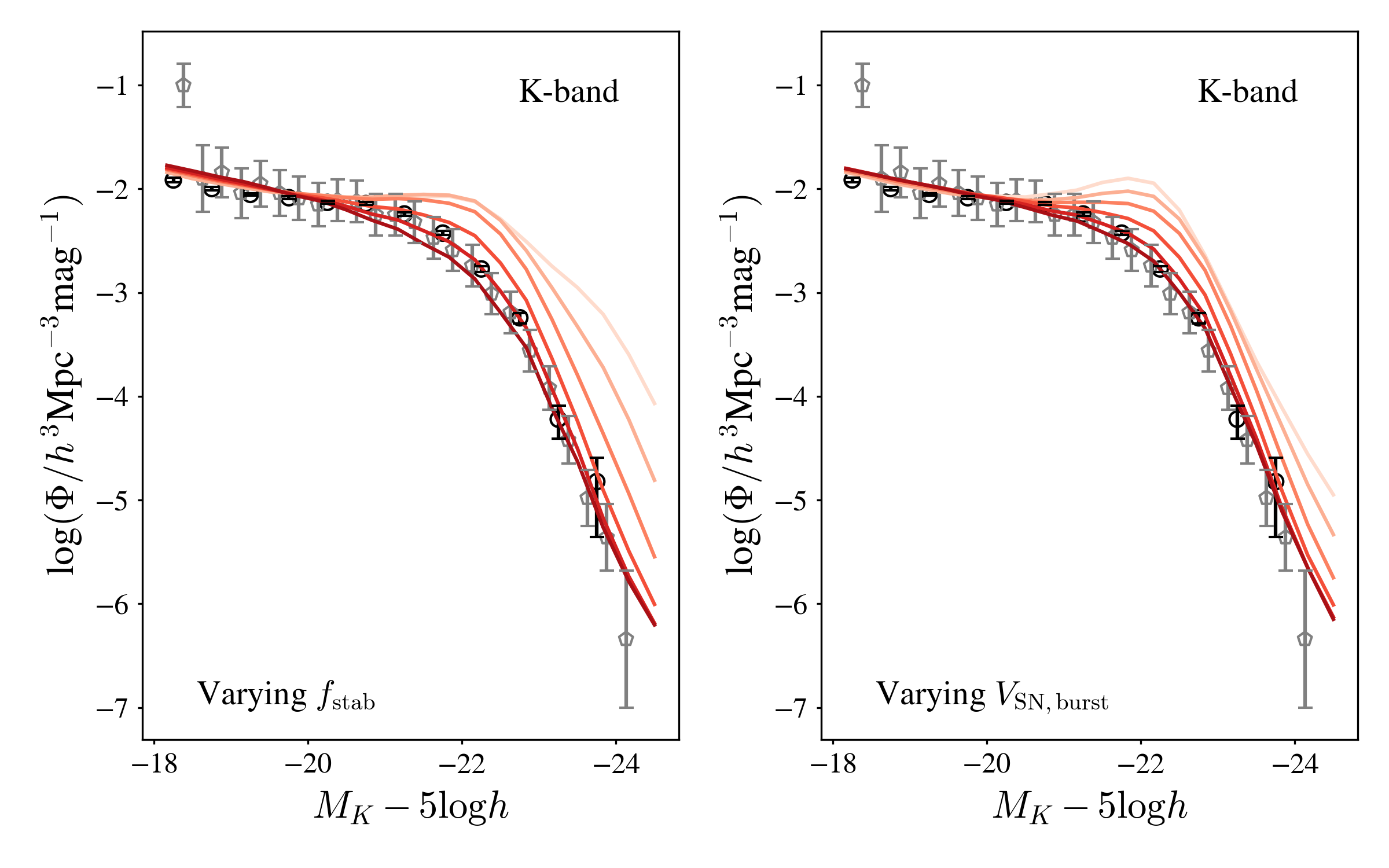}
    \caption{Emulator predictions for one-at-a-time perturbations of the parameters $f\textsubscript{stab}$ (\textit{left}) and $V\textsubscript{SN, burst}$ (\textit{right}) around a fit to the K-band luminosity function. We vary the parameters between the full range given in Table \ref{tab:ParamRanges}. Darker colours correspond to higher values. The data shown correspond to those described in \S\ref{sec:calibrationDatasetsDescription}.}
    \label{fig:OATVsnburst}
\end{figure}

Fig.~\ref{fig:OATVsnburst} illustrates the one-at-a-time effect of varying the parameters  $f\textsubscript{stab}$ and $V\textsubscript{SN, burst}$, as a demonstration of their degenerate effects.

\begin{figure}
    \centering
    \includegraphics[width=\columnwidth,trim={0.5cm 0.5cm 0cm 0cm},clip]{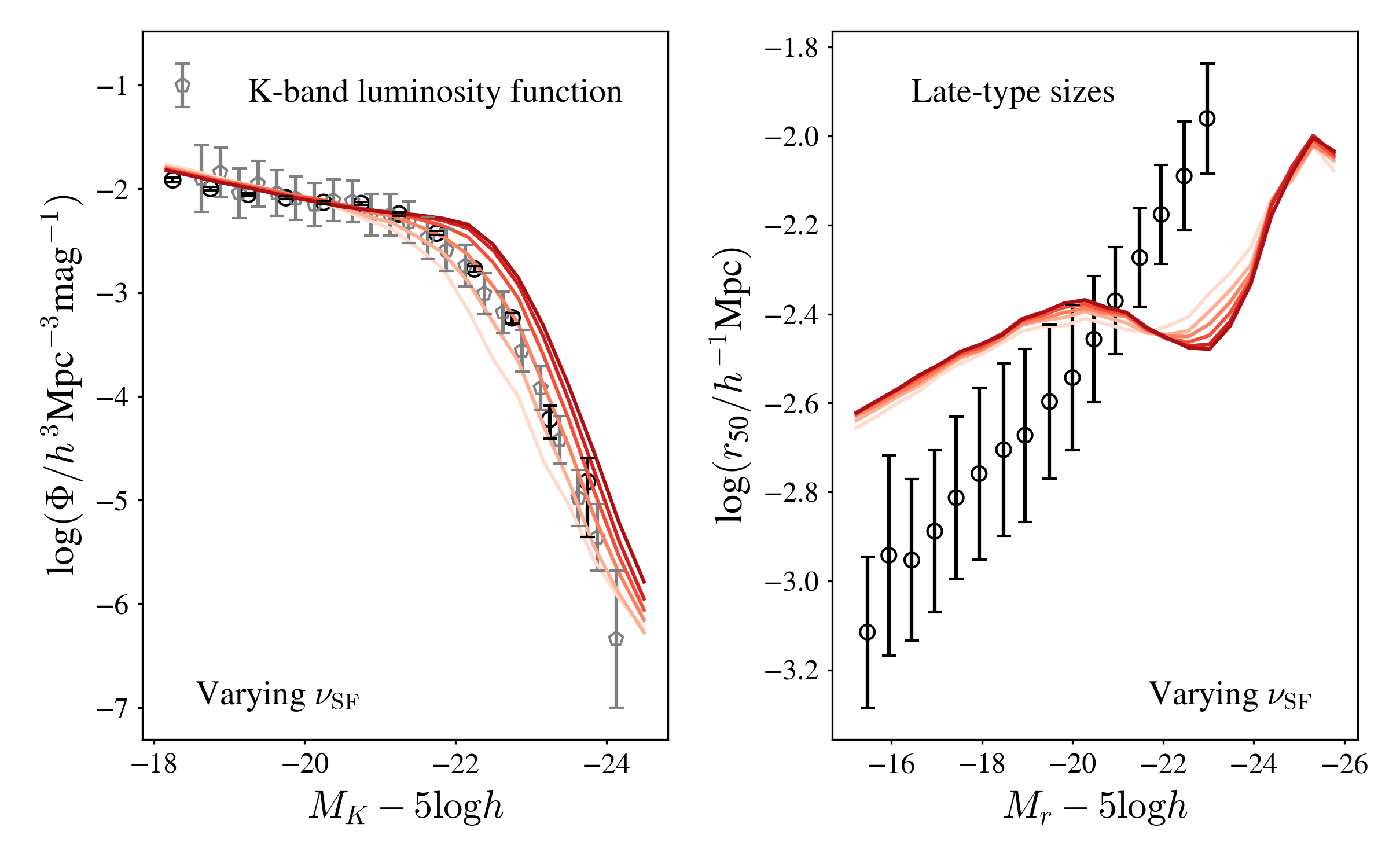}
    \caption{Emulator predictions for one-at-a-time perturbations of the parameter $\nu\textsubscript{SF}$ for the K-band luminosity function (\textit{left}) and the late-type galaxy sizes (\textit{right}) around a fit to the K-band luminosity function. We vary the parameters between the full range given in Table \ref{tab:ParamRanges}. Darker colours correspond to higher values. The data shown correspond to those described in \S\ref{sec:calibrationDatasetsDescription}.}
    \label{fig:OATnusf}
\end{figure}

Fig.~\ref{fig:OATnusf} shows the one-at-a-time effect of varying  $\nu\textsubscript{SF}$ on the K-band LF and late-type galaxy sizes. When fitting both the K-band LF and the late-type galaxy sizes, we see a decrease in the preferred value of $\nu\textsubscript{SF}$; Fig.~\ref{fig:OATnusf} demonstrates that this is because a lower $\nu\textsubscript{SF}$ counteracts the enhancement in the bright-end of the K-band LF caused by the higher value of $V\textsubscript{SN, disk}$ when including both constraints. We also see that reducing $\nu\textsubscript{SF}$ marginally improves the fit to the late-type galaxy sizes.

\begin{figure*}
    \centering
    \includegraphics[width=\textwidth,trim={3cm 3.5cm 3.5cm 3.8cm},clip]{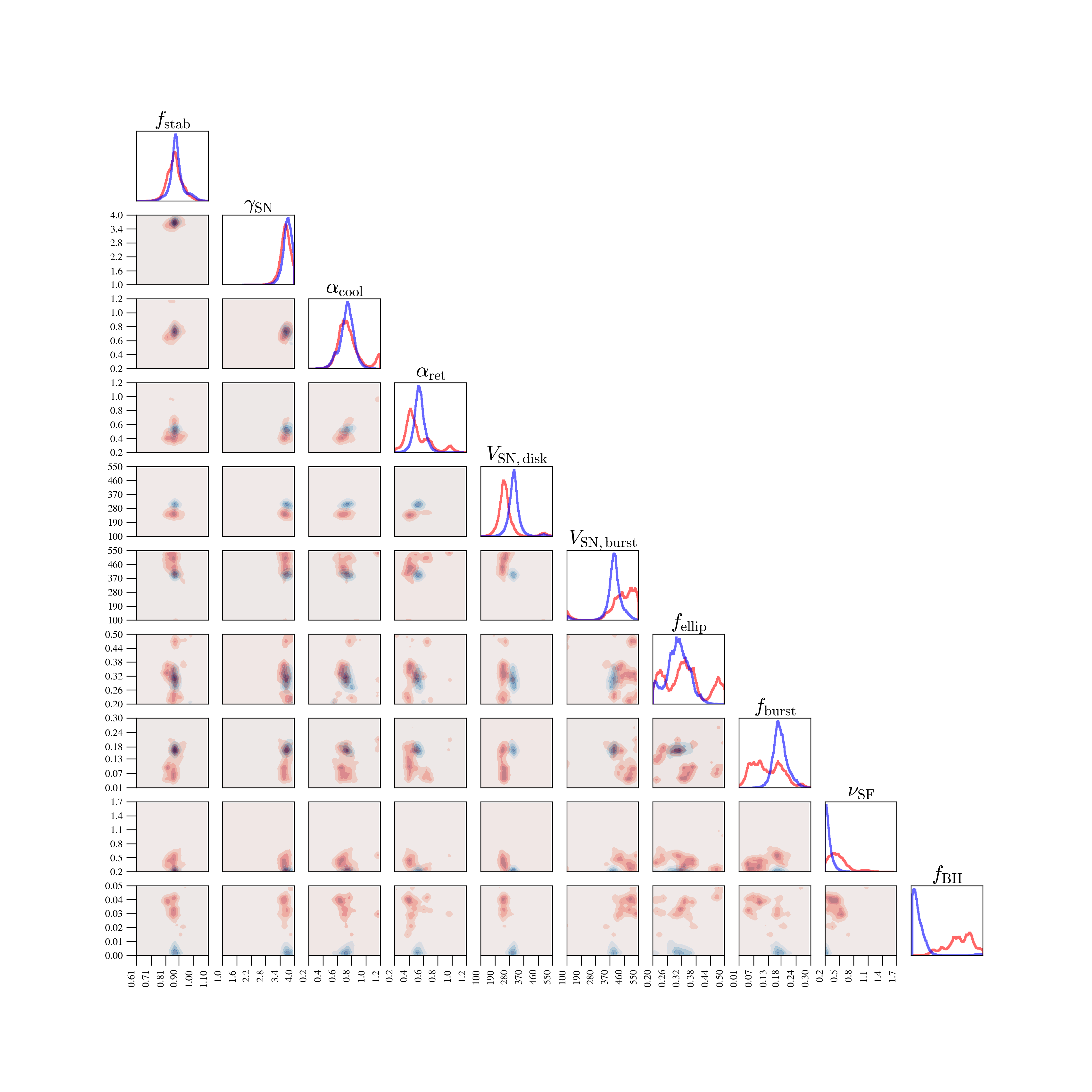}
    \caption{Accepted samples from 20 MCMC chains for fits to the K-band LF (\textit{red}), and both the K-band LF and the HI mass function (\textit{blue}). The first 50\% of samples were discarded to allow for burn-in. The histograms show the distribution of the parameters in 1D projection. The ranges on each axis are the same as those quoted in Table \ref{tab:ParamRanges}. The shading gives a sense of the density, with darker colours corresponding to more densely sampled regions. The darkest regions correspond to the 25th percentile, and the lighter regions to the 50th and 75th percentiles.}
    \label{fig:HItensionparams}
\end{figure*}

Fig.~\ref{fig:HItensionparams} shows the accepted parameters of 20 MCMC chains when we fit the K-band LF (red), and when we fit to the K-band LF and the HI mass function (blue). Here, we see that including the HI mass function results in higher values of $V\textsubscript{SN, disk}$ being preferred. $\nu\textsubscript{SF}$ is also moved to the bottom end of the explored range, and $\alpha\textsubscript{ret}$ becomes more sharply peaked and takes slightly higher values.

\bsp	
\label{lastpage}
\end{document}